\documentclass[apj]{emulateapj}

\newcommand {\HI}     {\ion{H}{1}}   
\newcommand {\HII}    {\ion{H}{2}}   
\newcommand {\OI}     {\ion{O}{1}}   
\newcommand {\SII}    {\ion{S}{2}}   
\newcommand {\SiII}   {\ion{Si}{2}}  
\newcommand {\SiIII}  {\ion{Si}{3}}  
\newcommand {\SiIV}   {\ion{Si}{4}}  
\newcommand {\OVI}    {\ion{O}{6}}   
\newcommand {\OVII}   {\ion{O}{7}}   
\newcommand {\OVIII}  {\ion{O}{8}}   

\newcommand {\CII}    {\ion{C}{2}}   
\newcommand {\CIII}   {\ion{C}{3}}   
\newcommand {\CIV}    {\ion{C}{4}}   
\newcommand {\CVI}    {\ion{C}{6}}   
\newcommand {\NV}     {\ion{N}{5}}   
\newcommand {\NVI}    {\ion{N}{6}}   
\newcommand {\NVII}   {\ion{N}{7}}   
\newcommand {\CaII}   {\ion{Ca}{2}}  
\newcommand {\NaI}    {\ion{Na}{1}}  

\newcommand {\kms}    {km~s$^{-1}$}
\newcommand {\etal}   {et~al.}
\newcommand {\cm}     {cm$^{-2}$}

\newcommand {\NSiIV}  {$N_{\rm SiIV}$}
\newcommand {\NSiII}  {$N_{\rm SiII}$}
\newcommand {\NSiIII} {$N_{\rm SiIII}$}
\newcommand {\NHI}    {$N_{\rm HI}$}

\newcommand {\FUSE}   {{\it FUSE}}
\newcommand {\HST}   {{\it HST}}
\newcommand {\IUE}   {{\it IUE}}

\begin{document}

\title{A Large Reservoir of Ionized Gas in the Galactic Halo:  Ionized Silicon in High-Velocity and Intermediate-Velocity Clouds}

\author{J. Michael Shull, Jennifer R. Jones\altaffilmark{1}, Charles W. Danforth, 
   \& Joseph A. Collins\altaffilmark{2}}
\affil{CASA, Department of Astrophysical and Planetary Sciences, \\
University of Colorado, 389-UCB, Boulder, CO 80309 \\
michael.shull@colorado.edu, charles.danforth@colorado.edu, jcollins@casa.colorado.edu}
\altaffiltext{1}{Now at Department of Physics \& Astronomy, Michigan State 
University, East Lansing, MI 48824, westbr39@msu.edu} 
\altaffiltext{2}{Also at Front Range Community College, Larimer Campus,
4616 S. Shields St., Fort Collins, CO 80526}



\begin{abstract}

The low Galactic halo is enveloped by a sheath of ionized,
low-metallicity gas, which can provide a substantial
($1~M_{\odot}$~yr$^{-1}$) cooling inflow to replenish star formation
in the disk.  Using absorption spectra from the {\it Hubble Space
Telescope} and {\it Far Ultraviolet Spectroscopic Explorer} toward 37
active galactic nuclei at high latitude, we detect widespread
interstellar \SiIII\ $\lambda1206.50$ absorption: 61 high-velocity
clouds (HVCs) along 30 sight lines and 22 intermediate-velocity clouds
(IVCs) along 20 sight lines. We find a segregation of redshifted and
blueshifted absorbers across the Galactic rotation axis at $\ell =
180^{\circ}$, consistent with a lag in the rotation velocity above the
Galactic plane.  The HVC sky coverage is large ($81 \pm 5$\% for 30
out of 37 directions) with \SiIII\ optical depth typically 4--5 times
that of \OVI\ $\lambda 1031.926$.  The mean HVC column density per
sight line, $\langle \log{\rm N}_{\rm SiIII} \rangle = 13.42 \pm
0.21$, corresponds to total column density N$_{\rm HII} \approx
(6\times10^{18}$~\cm)($Z_{\rm Si}/0.2Z_{\odot})^{-1}$ of ionized
low-metallicity gas, similar to that inferred in \OVI.  This reservoir
could total $10^8~M_{\odot}$ and produce a mass infall rate
$\sim1~M_{\odot}$~yr$^{-1}$.  If \SiII, \SiIII, \SiIV, and \HI\ are 
co-spatial and in photoionization equilibrium, the mean photoionization
parameter in the low halo, $\langle \log\,U \rangle = -3.0^{+0.3}_{-0.4}$, 
approximately ten times lower than observed in the low-redshift 
intergalactic medium.  The metallicities in some HVCs, derived 
from [\SiII/\HI], are 10--30\% solar, whereas values found from all 
three silicon ions are lower in the pure-photoionization models.  These 
formally lower metallicities are highly uncertain, since some of   
the higher ions may be collisionally ionized.  The HVC and IVC 
metallicities may be compared with the median photometric metallicity, 
[Fe/H] = $-1.46\pm0.30$, for $\sim$200,000 halo F/G stars in the 
Sloan Digital Sky Survey.

\end{abstract}


\section{Introduction}

Interstellar gas above the plane of the Milky Way disk has been studied
through a number of techniques, including emission in \HI\ (21~cm), 
soft X-rays, and diffuse H$\alpha$, pulsar dispersion measures, and
absorption lines in the optical, ultraviolet, and X-ray. These observations 
suggest a multiphase gaseous medium in the low halo, consisting of 
high-latitude absorption-line clouds, pressure-confined by a hot (coronal) 
gaseous background.  One of the first hints of this hot halo came from the 
detection of high-latitude interstellar clouds in 21-cm emission (Oort 1966)
and in \NaI\ and \CaII\ absorption (Adams 1949; M\"unch \& Zirin 1961; 
Hobbs 1965).  In order to provide for cloud confinement, Spitzer (1956) 
proposed the existence of a hot, ionized ``Galactic corona'' of 8 kpc 
vertical extent, with temperature $T \approx 10^6$~K and hydrogen 
density $n_H \approx 5 \times 10^{-4}$ cm$^{-3}$.

With the advent of ultraviolet spectrographs aboard the {\it
International Ultraviolet Explorer} (\IUE), the {\it Hubble Space
Telescope} (\HST), and the {\it Far Ultraviolet Spectroscopic
Explorer} (\FUSE), astronomers gained access to sensitive diagnostics
of hot halo gas, particularly the far-UV resonance lines of
high-ionization species such as \CIV, \SiIV, \NV, and \OVI\ (Savage \&
Sembach 1991; Shull \& Slavin 1994; Savage \etal\ 2003; Indebetouw \&
Shull 2004; Collins, Shull, \& Giroux 2003, 2007; Fox, Savage, \&
Wakker 2006).  Recent low-resolution spectra of active galactic nuclei
(AGN) taken with {\it Chandra} and {\it XMM-Newton} telescopes have
detected X-ray absorption lines at redshifts $z \approx 0$ (Nicastro
\etal\ 2002; Fang \etal\ 2002; McKernan \etal\ 2004) that may arise
from hot gas in the Galactic halo or Local Group.  These X-ray
absorbers include lines from \OVII\ (21.602~\AA), \OVIII\
(18.969~\AA), \NVI\ (28.787~\AA), \NVII\ (24.782~\AA), and \CVI\
(33.776~\AA).  Owing to the low resolution of the X-ray spectrographs,
the location of these ``$z=0$ absorbers'' remains controversial,
possibly residing in a Galactic halo of radial extent $r \sim 20$~kpc
(Bregman \& Lloyd-Davies 2007) or in a thick gaseous disk, of vertical
scale height $h \sim 2$ kpc (Yao \& Wang 2007; Yao \etal\ 2008).
  
In UV/optical spectra, the observed ion stages in high-latitude clouds
range from low-ionization states (\HI, \OI, \SII, \SiII, \NaI, \CaII)
to ionized gas traced by \CIII, \CIV, \NV, \OVI, \SiIII, and \SiIV.
These ``high ions'' could arise from both photoionization and
collisional ionization, depending on cloud proximity to sources of
ionizing photons or interfaces with hot gas. They are best traced by
absorption in the ultraviolet lines of \OVI\ ($\lambda 1031.926,
1037.617$), \NV\ ($\lambda1238.821, 1242.802$), and \CIV\ ($\lambda
1548.195, 1550.770$).  Ultraviolet surveys of the vertical
distribution of these high ions suggest a layer of hot gas, with scale
heights ranging from 3--5 kpc (Sembach \& Savage 1992; Shull \& Slavin
1994).  This corona is expected to lie near the virial temperature,
$T_{\rm vir} \approx (2 \times 10^6~{\rm K}) \, M_{12}^{2/3}
\left[ (1+z_{\rm vir}) / 4 \right]$ for the Milky Way, with total mass 
$M = (10^{12}~M_{\odot}) M_{12}$ assembled at virialization redshift 
$z_{\rm vir} \approx 3$.  Laced with heavy elements, this gas is expected to 
cool into clumps (Maller \& Bullock 2004; Collins, Shull, \& Giroux 2004), 
fall inward, and intercept uprising gas from superbubbles in the Galactic disk.  
The result is a disk-halo interface and Galactic fountain 
(Shapiro \& Field 1976; Bregman 1980; Houck \& Bregman 1990; Shull \& Slavin 1994).  
It has been suggested that \OVI\ and \NV\ may reside in bow shocks and other 
interfaces between infalling clouds and the hot gaseous corona 
(Indebetouw \& Shull 2004; Fox \etal\ 2006).

In many disk galaxies, the infall of low-metallicity gas seems to be
important for gas replenishment, the evolution of star formation
rates, and the chemical imprint on the mass-metallicity relations
(Tremonti \etal\ 2004; Erb \etal\ 2006).  In the Milky Way, the
processes of gaseous infall and recycling are observed in Galactic
high-velocity clouds (HVCs) and intermediate velocity clouds (IVCs),
defined as clouds of primarily neutral hydrogen whose velocities are
inconsistent with the standard model of Galactic rotation.  The
distinction between HVCs and IVCs lies in their local-standard-of-rest
(LSR) velocities, with a nominal population break at $|V_{\rm LSR}| =
90-100$ \kms.  Following the convention of Wakker \& van Woerden
(2007), we define IVCs to lie between $|v_{\rm LSR}| \approx$
30--90\,\kms\ and HVCs to have $|v_{\rm LSR}| > 90$~\kms.  By
analyzing the metallicity of this low-halo gas, we can constrain their
spatial location and origin.  Galactic fountains are expected to be
metal-rich, so the resulting absorbers should have metallicities near
solar.  By contrast, gas falling into the disk from the Local Group or
intergalactic medium (IGM) would be metal-poor, perhaps 10--30\% of
solar values.

Two interpretations have arisen for the HVC and IVC positions in the
Galaxy.  The first model places HVCs in the low Galactic halo (Collins
\etal\ 2003; Tripp \etal\ 2003; Fox \etal\ 2006) as inflows onto and
outflows from the disk.  The second model identifies a sub-class of
compact HVCs, at locations in the Local Group (Blitz \etal\ 1999).
Constraining the ionization conditions may distinguish whether HVCs
are similar to gas from the Galactic disk, within the Local Group, or
in the IGM.  This uncertainty in HVC distances to HVCs creates a
semantic problem in defining their location in the ``thick disk''
(scale height $h \leq 1$ kpc) or ``low halo'' ($h \approx$ 5--10 kpc).
Distance measurements have recently been reported for several HVCs,
using the presence or absence of \CaII\ absorption toward Galactic halo
stars.  These studies bracket the distance to HVC Complex~C at $10 \pm
2.5$ kpc (Thom \etal\ 2008; Wakker \etal\ 2007) and place other HVCs
distances at $\sim10$ kpc (Wakker \etal\ 2008).  Complex~C is a good
example of infalling metal-poor gas (Wakker \etal\ 1999; Collins
\etal\ 2003, 2007; Tripp \etal\ 2003).  By analyzing [\OI/\HI] along
11 AGN sight lines through Complex~C, Collins \etal\ (2007) find a
column-density weighted metallicity of $0.13 Z_{\odot}$.  Toward
PG\,1259$+$593, another AGN sight line through Complex~C, the IVC has a
near-solar [\OI/\HI] abundance, while the HVC has an abundance
$\sim10$\% solar (Richter \etal\ 2001; Sembach \etal\ 2004; Collins
\etal\ 2007).  These observations suggest that IVCs are material near
the Galactic plane, while HVCs have fallen in from more distant gas.

Many groups have used data from \HST, \FUSE, and the Leiden/Dwingeloo 
Survey (LDS) to derive column densities of a variety of interstellar HVC 
species.  Savage \etal\ (2003) studied the distribution and kinematics of 
\OVI\ in the Galactic halo, using \FUSE\ spectra of \OVI\ over the velocity range
$-100 < v_{\rm LSR} < +100$ \kms, along paths toward 100 AGN and two
distant halo stars.  Sembach \etal\ (2003) studied \OVI\ HVCs in the
same 102 sight lines, identifying 84 high-velocity features at $-500 <
v_{\rm LSR} < +500$ \kms, with mean column density $\langle \log
N_{\rm OVI} \rangle = 13.95 \pm 0.34$.  These \OVI\ absorbers exhibit
velocity centroids ranging from $-372$ to $+385$ \kms\ and line
widths, expressed as doppler velocities, of $b_{\rm OVI} =$ 16--72
\kms. High-velocity \OVI\ was seen along 59 of 102 sight lines,
suggesting that $\sim 60$\% of the sky, and perhaps even higher
controlling for data-quality, is covered by high-velocity ionized gas.
This covering fraction is higher than that seen in 21-cm emission,
measured at 37\% for column densities \NHI\ $\geq 7\times10^{17}$~\cm\
(Murphy, Lockman, \& Savage 1995).

In this paper, we focus on silicon in three ionization 
states accessible to UV spectroscopy with \HST\ and \FUSE.  
We analyze multiple strong ultraviolet absorption features, including 
\SiII\ (1020.70, 1190.42, 1193.29, 1260.42, 1304.37,
1526.71~\AA), \SiIII\ (1206.50~\AA), and \SiIV\ (1393.76,
1402.77~\AA).  As background targets (Figure~1) we chose 37 AGN
distributed over high Galactic latitudes, $|b|\geq20\arcdeg$, to avoid
confusion from absorption in the Galactic plane.  Our previous
spectroscopic studies of HVCs (Gibson \etal\ 2000, 2001; Collins
\etal\ 2003, 2007) compared ions of many elements (C, N, O, Si, S, Fe)
in Complex~C and the Magellanic Stream, although
[\OI/\HI] is the best indicator of metallicity, owing to
charge-exchange coupling. Assumptions about the relative metallicity
between elements (O, Si, S, Fe) can be bypassed by observing multiple
ionization states of the same element.  In the ultraviolet, multiple
ions are available for silicon (Si\,II/III/IV), as well as carbon
(C\,II/III/IV), sulfur (S\,II/III/IV), and iron (Fe\,II/III).

This paper is organized as follows.  In \S~2, we discuss the importance
of the strong \SiIII\ $\lambda1206.50$ absorption line as a probe of
Galactic HVCs and IVCs.  We then describe the acquisition and reduction 
of the 21-cm data from the LDS, and the ultraviolet absorbers (\SiII, \SiIII, 
\SiIV) with \FUSE\ and \HST.
Our survey found 61 HVCs and 22 IVCs along 37 AGN sight lines. 
In \S~3, we discuss our data analysis and photoionization modeling.  
In \S~4 we discuss our results, placing these observations in
the context of other measures of the Galactic halo and low-redshift IGM.


\begin{figure*}
\epsscale{1}
\plotone{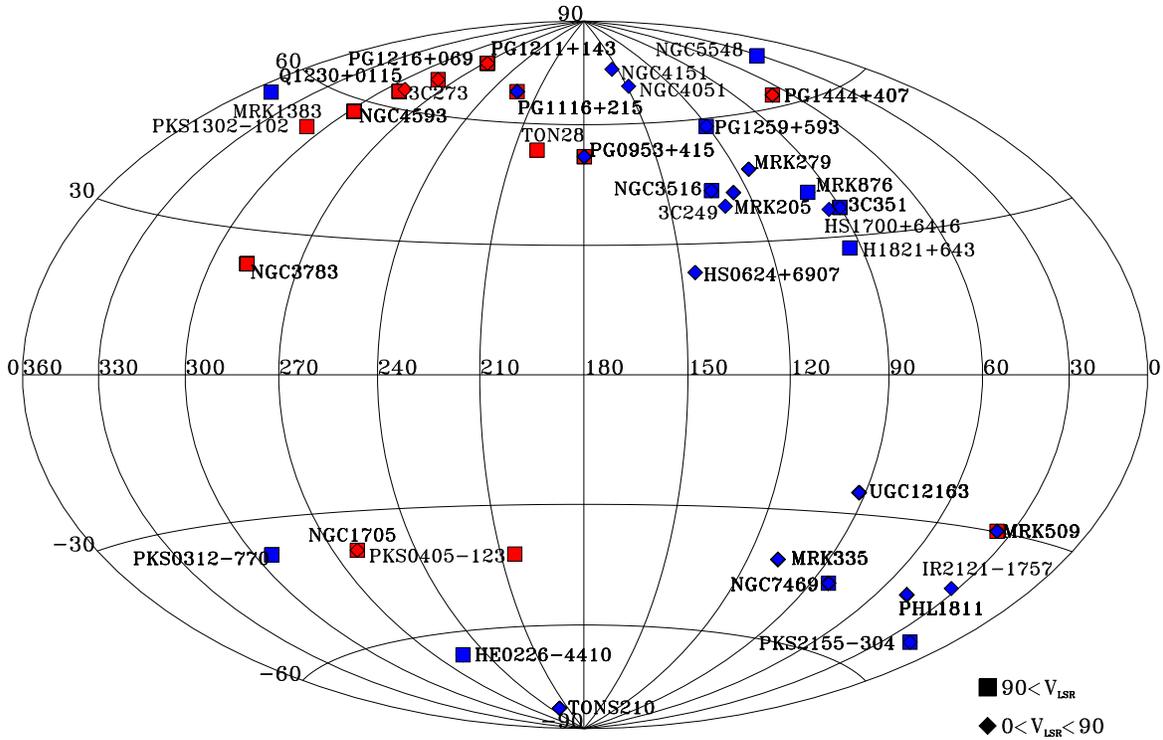}
  \caption{Positions of 37 observed AGN sight lines (25 in northern
   Galactic hemisphere and 12 in south) plotted in Galactic Coordinates,
   with anticenter ($\ell = 180^{\circ}$) shown in the middle.  Central
   velocities of HVC and IVC absorption features in each sight line are
   color-coded as shown in the legend.  Of the 37 sight lines, 30 (81\%)
   showed high-velocity \SiIII. Multiple HVCs were seen along 20 sight
   lines (8 with three or more HVCs), while 7 sight lines showed no
   HVCs in \SiIII\ (only an IVC).  The HVC redshifts and blueshifts are
   largely segregated on either side of the Galactic rotation axis through
   $\ell = 180^{\circ}$. See text for possible interpretation as a dropoff
   in circular velocity above the disk. }
\end{figure*}



\section{Silicon Ions in HVCs and IVCs}

\subsection{The Importance of Si\,III} 

We begin with the strong \SiIII\ $\lambda1206.500$ absorption line, for which
we have recently completed an HVC survey (Collins, Shull, \& Giroux 2009).
This line has a large oscillator strength, $f = 1.63$ (Morton 2003), making it
a good tracer of ionized gas, comparable in sensitivity to the \OVI\ doublet
(1031.926, 1037.617~\AA).  The \SiIII\ absorption line is intrinsically 14 times 
stronger (in $f \lambda$) than
\OVI\ $\lambda 1031.926$, which compensates for the lower abundance ratio,
(Si/O)$_{\odot} \approx 0.071$ based on solar abundances,
(Si/H)$_{\odot} = 3.24 \times 10^{-5}$ and (O/H)$_{\odot} = 4.57 \times 10^{-4}$,
from Asplund \etal\ (2005). With STIS/E140M, the \SiIII\ line 
provides sensitivity to column densities at or below 
N$_{\rm SiIII} \approx (10^{12}$~\cm)$N_{12}$, corresponding to equivalent 
width $W_{\lambda} =$ (21.0~m\AA)$N_{12}$.  For a Gaussian profile,
the \SiIII\ optical depth at line center is 
$\tau_0 \approx (0.295) N_{12} b_{10}^{-1}$, scaled to a doppler width 
$b_{\rm SiIII}$ = (10~\kms)$b_{10}$, chosen to reflect the turbulence and 
velocity components that broaden HVC metal-line absorbers (Collins \etal\ 2003, 
2007).  Typical HVCs exhibit \SiIII\ absorption over an extended velocity range, 
$\Delta v = 40-100$ \kms, far broader than any single component would allow.  
Indeed, the curves of growth derived for low metal ions in HVC Complex C 
(Collins \etal\ 2003, 2007) range from $b \approx$ 7--18 \kms, consistent 
with non-thermal line widths.  Given these large velocity dispersions, we 
expect \SiIII\ line saturation to set in at column densities 
N$_{\rm SiIII} > (3.4 \times 10^{12}~{\rm cm}^{-2})b_{10}$ in each 
STIS/E140M resolution element.  In our column-density tables, we express 
log \NSiIII\ as a lower limit when saturation is believed to exist.

\begin{deluxetable*}{lccrrrr}[t]
\tabletypesize{\scriptsize}
\tablecolumns{7}
\tablewidth{0pt}
\tablecaption{Total AGN Sight Lines in Survey}
\tablehead{\colhead{Sightline}                                  &
          \colhead{RA\,(J2000)}                                 &
          \colhead{Dec\,(J2000)}                                &
          \colhead{$\ell$}                                      &
          \colhead{$b$}                                         &
          \colhead{\FUSE}                                       &
          \colhead{\HST}                                        \\
          \colhead{}                                            &
          \colhead{({h}\phn{m}\phn{s})}                         &
          \colhead{(\phn{\arcdeg}~\phn{\arcmin}~\phn{\arcsec})} &
          \colhead{(\phn{\arcdeg})}                             &
          \colhead{(\phn{\arcdeg})}                             &
          \colhead{(ksec)}                                      &
          \colhead{(ksec)}                                      }
\startdata
Mrk\,335            & 00 06 19.5 & $+$20 12 10.3 & 108.76 & $-$41.42 & 97.0    & 17.1\\
Ton\,S210           & 01 21 51.6 & $-$28 20 57.3 & 224.97 & $-$83.16 & 52.2    & 27.4\\
HE\,0226$-$440      & 02 28 15.2 & $-$40 57 16.0 & 253.9  & $-$65.77 & 64.2    & 3.02\\  
PKS\,0312$-$770     & 03 11 55.4 & $-$76 51 50.8 & 293.4  & $-$37.55 & 15.4    & 12.6\\  
PKS\,0405$-$123     & 04 07 48.2 & $-$12 11 31.5 & 204.93 & $-$41.76 & 71.1    & 27.4\\
NGC\,1705           & 04 54 13.5 & $-$53 21 39.8 & 261.0  & $-$38.74 & 15.4    & 3.11\\  
HS\,0624$+$6907     & 06 30 02.7 & $+$69 05 03.7 & 145.71 & 23.35    & 112.3   & 62.0\\
PG\,0953$+$415      & 09 56 52.4 & $+$41 15 22.0 & 179.79 & 51.71    & 72.1    & 27.4\\
Ton\,28             & 10 04 02.6 & $+$28 55 35.5 & 200.08 & 53.21    & 11.2    & 33.0\\
3C\,249             & 11 04 13.7 & $+$76 58 58.2 & 130.39 & 38.55    & 216.8   & 68.8\\
NGC\,3516           & 11 06 47.6 & $+$72 34 06.9 & 133.24 & 42.40    & 75.1    & 20.6\\
PG\,1116$+$215      & 11 19 08.7 & $+$21 19 18.2 & 223.36 & 68.21    & 77.0    & 26.5\\
NGC\,3783           & 11 39 01.7 & $-$37 44 18.9 & 287.4  & 22.94    & 37.3    & 2.7 \\  
NGC\,4051           & 12 03 09.6 & $+$44 31 52.8 & 148.8  & 70.08    & 28.8    & 2.2 \\  
NGC\,4151           & 12 10 32.6 & $+$39 24 21.0 & 155.08 & 75.06    & 97.8    & 13.2\\
PG\,1211$+$143      & 12 14 17.6 & $+$14 03 12.7 & 267.55 & 74.31    & 52.3    & 42.5\\
PG\,1216$+$069      & 12 19 20.9 & $+$06 38 38.4 & 281.0  & 68.14    & 12.4    & 2.90\\     
Mrk\,205            & 12 21 44.0 & $+$75 18 38.3 & 125.45 & 41.67    & 203.6   & 62.1\\
3C\,273             & 12 29 06.7 & $+$02 03 08.9 & 289.95 & 64.36    & 42.3    & 18.7\\
Q\,1230$+$0115      & 12 30 50.0 & $+$01 15 21.7 & 291.26 & 63.66    & 4.0     & 9.8 \\
NGC\,4593           & 12 39 39.5 & $-$05 20 38.0 & 297.48 & 57.40    & \nodata & 11.0\\
PG\,1259$+$593      & 13 01 13.1 & $+$59 02 05.7 & 120.56 & 58.05    & 668.3   & 95.8\\
PKS\,1302$-$102     & 13 05 33.0 & $-$10 33 19.3 & 308.5  & 52.16    & 83.3    & 11.0\\  
Mrk\,279            & 13 53 03.4 & $+$69 18 29.9 & 115.04 & 46.86    & 228.5   & 54.6\\
NGC\,5548           & 14 17 59.5 & $+$25 08 12.4 & 31.96  & 70.50    & 55.0    & 69.8\\
Mrk\,1383           & 14 29 06.6 & $+$01 17 06.6 & 349.22 & 55.13    & 63.5    & 10.5\\
PG\,1444$+$407      & 14 46 45.9 & $+$40 35 06.4 & 69.90  & 62.72    & 10.0    & 48.6\\
Mrk\,876            & 16 13 57.2 & $+$65 43 09.6 & 98.27  & 40.38    & 46.0    & 29.2\\ 
HS\,1700$+$6416     & 17 01 00.6 & $+$64 12 09.9 & 94.40  & 36.16    & 285.5   & 9.1 \\
3C\,351             & 17 04 41.5 & $+$60 44 28.0 & 90.08  & 36.38    & 141.9   & 77.0\\
H\,1821$+$643       & 18 21 57.3 & $+$64 20 36.3 & 94.00  & 27.42    & 132.3   & 50.9\\
Mrk\,509            & 20 44 09.7 & $-$10 43 24.7 & 35.97  & $-$29.86 & 62.3    & 7.6 \\
IR\,2121$-$1757     & 21 24 41.7 & $-$17 44 45.9 & 32.78  & $-$41.64 & \nodata & 5.3 \\
PHL\,1811           & 21 55 01.6 & $-$09 22 26.0 & 47.47  & $-$44.82 & 75.0    & 33.9\\
PKS\,2155$-$304     & 21 58 52.1 & $-$30 13 32.3 & 17.73  & $-$52.25 & 123.2   & 10.8\\
UGC\,12163          & 22 42 39.3 & $+$29 43 31.3 & 92.14  & $-$25.34 & 60.9    & 10.3\\
NGC\,7469           & 23 03 15.6 & $+$08 52 26.2 & 83.10  & $-$45.47 & 44.3    & 22.8\\
\enddata
\end{deluxetable*}

To compare the \SiIII\ and \OVI\ observations, it is useful to estimate 
the expected relative absorption-line strengths. The absorption optical 
depth at frequency $\nu$ scales as 
$\tau_{\nu} \propto N f \lambda / (\Delta v)$, where $N$ is the column
density, $f \lambda$ measures the intrinsic line strength, and $\Delta v$ 
is the effective line width of the absorption profile.  The ratio
of optical depths of \SiIII\ $\lambda1206.50$ and \OVI\ $\lambda1031.93$ 
is then,
\begin{equation}
  \frac {\tau_{\rm SiIII}} {\tau_{\rm OVI}} = (1.02) \left[ \frac
      {Z_{\rm Si} \, f_{\rm SiIII} \, {\rm N}_{\rm H}^{\rm SiIII} \Delta v_{\rm OVI}}
      {Z_{\rm O} \, f_{\rm OVI} \, {\rm N}_{\rm H}^{\rm OVI} \Delta v_{\rm SiIII}} 
       \right]  \; ,  
\end{equation}
where $Z_{\rm Si}$ and $Z_{\rm O}$ are silicon and oxygen metallicities, 
$f_{\rm SiIII}$ and $f_{\rm OVI}$ are the ionization fractions of these 
species, $\Delta v_{\rm SiIII}$ and $\Delta v_{\rm OVI}$ 
are the effective line widths of the two lines, and
N$_{\rm H}^{\rm SiIII}$ and N$_{\rm H}^{\rm OVI}$ are the hydrogen column 
densities of the multiphase gas traced by \SiIII\ and \OVI\, respectively.  
Because the line widths are probably non-thermal, we expect that 
$\Delta v_{\rm OVI} / \Delta v_{\rm SiIII} \approx 1$, rather than the
higher values suggested by their mass-dependent thermal doppler widths,  
$\Delta v_{\rm OVI}/\Delta v_{\rm SiIII} = 1.32(T_{\rm OVI}/T_{\rm SiIII})^{1/2}$. 
The ionization fractions, $f_{\rm SiIII} \approx 0.7\pm0.2$ and 
$f_{\rm OVI} \approx 0.2\pm0.1$, are also uncertain and depend 
on the relative importance of photoionization and collisional ionization. 
In the multiphase conditions likely in HVCs, these ionization fractions  
can have wide ranges. However, \SiIII\ (a Mg-like ion with a closed-subshell 
$3s^2$ valence structure) typically has a much larger ionization fraction 
than \OVI\ (a Li-like ion with a weakly bound $2s$ valence electron).  
For example, if \SiIII\ and \OVI\ each exist in collisional ionization 
equilibrium, their peak ionization fractions (Sutherland \& Dopita 
1993) would be $f_{\rm SiIII} = 0.90$ at $\log T_{\rm max} = 4.45$ and 
$f_{\rm OVI} = 0.22$ at $\log T_{\rm max} = 5.45$.  Therefore, adopting an 
ionization ratio $f_{\rm SiIII}/f_{\rm OVI} \approx 4$ and setting the
other scaling ratios in eq.\ (1) equal to unity, we expect \SiIII\ to be 
stronger than \OVI, with $\tau_{\rm SiIII} / \tau_{\rm OVI} \approx 4$.

The above estimate is merely illustrative, since \SiIII\ and \OVI\ probably 
reside in different spatial and thermal portions of the HVCs.  For example,  
if the HVC is an infalling clump of cooling gas from the halo, the \HI\ 
and low ions (\SiII, \SII, \OI) are likely to be a mixture of cool 
gas at $T < 10^3$~K and photoionized gas at $10^4$~K.  The higher ions
(\OVI, \NV) may come from bow shocks or interfaces with hotter
halo gas and have $T > 10^5$~K.  The intermediate ions (\SiIII, \SiIV, 
\CIII, \CIV) may arise in both photoionized and collisionally ionized gas,
complicating the ionization corrections needed to derive metallicities. 
Such environments include bow shocks around infalling HVCs, expanding shells 
around wind-driven bubbles, galactic winds, or supershells.  These 
multiphase effects, in which \SiIII\ and \OVI\ do not reside in the same 
gas, are captured in eq.\ (1) by the ratio, 
${\rm N}_{\rm H}^{\rm SiIII}/{\rm N}_{\rm H}^{\rm OVI}$.
  
These considerations suggest that \SiIII\ is a strong tracer of ionized gas, 
probing different thermal phases and ionization states than \OVI.  Our 
observations show that \SiIII\ $\lambda1206$ is typically 4--5 times 
stronger than \OVI\ $\lambda1032$, as illustrated in Figure~2 for 
eight sight lines with HVCs and IVCs.  Therefore, \SiIII\ provides an 
excellent probe of low column density (ionized) HVCs, which can be used 
with \OVI, \NV, \CIV, \SiIV, and other ions to trace the content and 
metallicity of ionized 
gas at high latitude.  One limitation to the single \SiIII\ line is the 
the possibility of line saturation. Without any second \SiIII\ transition, 
it is difficult to use strong $\lambda1206$ absorption systems, except to 
define the velocity range and provide ionization information in tandem with 
\SiII\ and \SiIV.  In our survey, we use \SiIII\ to identify weak HVCs
and rely on \SiII\ and \SiIV\ to separate HVCs into velocity components. 


\begin{figure*}
\epsscale{1}
\plotone{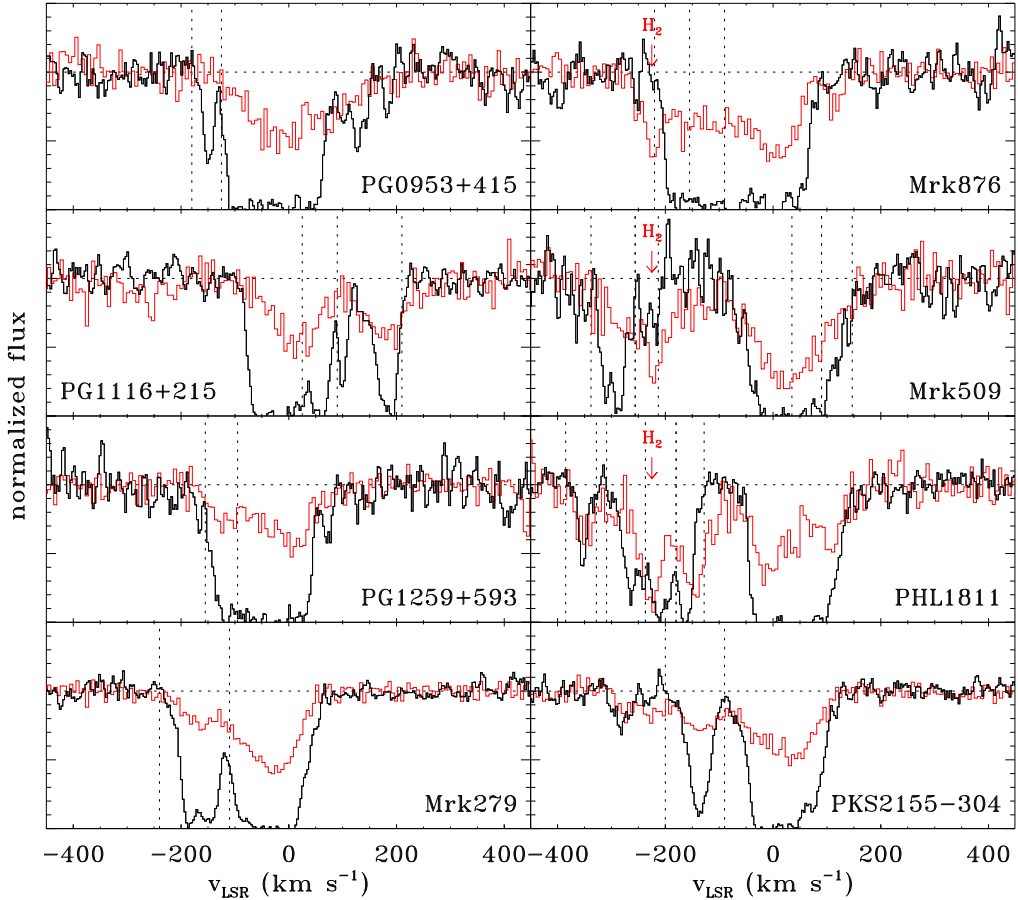}
  \caption{Ultraviolet spectra (\HST/STIS and \FUSE) of 8 AGN, showing
    absorption spectra of \SiIII\ $\lambda1206.50$ (black) and \OVI\
    $\lambda1031.93$ (red) vs.\ LSR velocity.  Velocity ranges
    for HVCs and IVCs are noted with
    vertical dashed lines, as is the position of H$_2$ absorption
    along 3 sight lines.  Note that \SiIII\ is typically 4--5 times stronger
    than \OVI\ in the HVCs, making it an excellent tracer of small
    amounts of ionized gas. }
\end{figure*}


In the next section, we focus on what can be learned assuming that the 
three silicon ions in HVCs and IVCs are predominantly photoionized.  
A significant advance in HVC photoionization modeling is enabled by access 
to three Si ionization stages, allowing us to constrain the
photoionization parameter, $U = n_{\gamma}/n_H$, where $n_{\gamma}$ is the
density of ionizing photons and $n_H$ is the density of hydrogen nuclei.
Combined with \HI\ column densities, \NHI, these parameters enable us
to derive physical models of the cloud densities ($n_H$), elemental 
abundances, and silicon metallicities, $Z_{\rm Si}$.
>From the distribution of ionization parameters and metallicities, we hope to
define the spatial location of HVCs and IVCs from the Galactic
plane and infer the nucleosynthetic sources of their heavy elements.

\subsection{AGN Sight Lines and their HVC/IVC Features}

>From our database of \HST\ and \FUSE\ spectra of high-latitude AGN
(Danforth \& Shull 2005, 2008), we selected 37 sight lines (see Figure~1)
to search for the presence of \SiII, \SiIII, and \SiIV.  These particular 
targets (Table~1) were selected from a larger sample of 58 AGN 
(Collins, Shull, \& Giroux 2009) observed with both the E140M echelle and 
G140M gratings.  These 37 sight lines were studied both for the presence 
of HVCs and IVCs and for the quality of their data.  
The minimum signal-to-noise per resolution element is 
(S/N)$_{\rm min} = 3$, and the mean is $\langle {\rm S/N \rangle} = 13$. 

The LDS data were analyzed for the \HI\ emission spectra and column densities, 
\NHI, as described in Wakker \etal\ (2003) and Tumlinson \etal\ (2004), having 
been reduced by Hartmann \& Burton (1997).   
At first, it appeared that the \HI\ data would give 
the best limits for the velocity range of HVCs and IVCs. However, in most cases, 
the ranges were better determined by silicon absorption features themselves.
The strong \SiIII\ $\lambda1206.5$ line is a sensitive tracer of
ionized gas, although often saturated.  In those cases, the weaker \SiII\ 
lines usually trace the velocity ranges of the cooler HVC components.  

For spectra with \SiII, \SiIII, and \SiIV\ absorption features, we used the
{\it Space Telescope Imaging Spectrograph} (STIS) echelle on \HST.
We analyzed STIS data for \SiIII\ and \SiIV\ lines in all 83 absorbers
(61 HVCs and 22 IVCs) and for \SiII\ lines.  In some HVC/IVCs, we also
analyzed \FUSE\ data for \SiII\ $\lambda1020.699$.  Some of these absorbers 
are ``highly ionized HVCs'', originally defined (Sembach \etal\ 1995, 1999) 
through the presence of \CIV\ or \OVI\ absorption, but with no \HI\   
21-cm emission.  For example, the sight lines to PKS\,2155$-$304
and Mrk\,509 exhibit several of this absorber type, in which the HVC is
detected in high ions (\CIV, \OVI) but not in \HI\ emission.
Many of these highly ionized HVCs are detected in \HI\ through their
Lyman-series absorption lines and in low ions such as \OI, \CII, \SiII, 
and \SII\ (Collins \etal\ 2004, 2005).  These absorbers are probably an 
extension of the HVC population to column densities, 
\NHI\ $\leq 10^{18}$~\cm, below the sensitivity level of 21-cm emission
surveys. 

The \FUSE\ data were retrieved from the Multimission Archive at the Space 
Telescope Science Institute (MAST) and reduced locally using {\sc calfuse}v2.4
\footnote{Detailed calibration information can be found at
{\tt http://fuse.pha.jhu.edu/analysis/calfuse\_intro.html}}.  Raw
exposures within a single \FUSE\ observation were coadded by channel
midway through the pipeline.  For further information on the reduction
method see Danforth \etal\ (2006).  The \FUSE\ data have spectral
resolution $\sim20$~\kms, while the STIS echelle (E140M) has a
resolution of $\sim7$~\kms.  (The \FUSE\ data were binned by three
pixels, where \FUSE\ resolution is typically 8-10 pixels.)  Archival
STIS/E140M echelle spectra were reduced locally by S. Penton and
smoothed over three pixels to form the 7 \kms\ resolution element.
The \FUSE\ and STIS data were both normalized in 10 \AA\ segments
centered on the rest wavelength of each absorption feature. The
continuum regions were selected interactively and fitted using
Legendre polynomials up to sixth order. The signal-to-noise ratio,
$S/N$, was designated as a 1$\sigma$ deviation from the fitted
continuum. However for the \FUSE\ data, we oversampled the resolution
element (15--20 \kms) by three pixels.  The actual signal-to-noise is
then $(S/N)_{\rm res}=\sqrt{3}\,(S/N)_{\rm pix}$.  The STIS data were
smoothed to the instrumental resolution prior to normalization, and
the velocity was converted to the LSR for consistency.

In the STIS band, we examined five absorption lines: 
\SiII\ $\lambda1304.370$, \SiII\ $\lambda1526.707$, \SiIII\ $\lambda1206.500$, 
and the \SiIV\ doublet, $\lambda\lambda 1393.755,\,1402.770$.  
We did not include several other 
\SiII\ lines ($\lambda$1190.4, 1193.3, 1260.4), because they were blended 
with other interstellar features or too saturated to be of value.  However, 
\SiII\ $\lambda1020.699$ lies in the \FUSE\ band, and in some cases 
produced a cleaner absorption feature.  As a weaker line, \SiII\ $\lambda1020$ 
often provided a better velocity constraint than the saturated \SiII\ lines 
seen with STIS. 

The color coding in Figure~1 illustrates the partial segregation of
redshifts and blueshifts on either side of the Galactic rotation axis
passing through $\ell = 180^{\circ}$.  This effect has been seen in
other species (Collins \etal\ 2005; Fox \etal\ 2006).  Because
HVC velocities are plotted in the LSR, the observed velocity segregation could
be the result of the falloff in circular rotation velocity with elevation
above the Galactic plane.  In defining the LSR velocity for high-latitude
HVCs, based on the Galactic rotation curve in the plane, we may have 
subtracted too large a circular velocity for directions
at $\ell < 180^{\circ}$, and the reverse at $\ell > 180^{\circ}$.  The
result would be an apparent asymmetry in LSR velocities across the
$\ell = 180^{\circ}$ axis, as seen generally in Fig.\ 1.
Although most previous HVC models assume that the circular velocity remains
constant on cylinders, as one moves above the Galactic plane, a velocity lag
has recently been noted in the Sloan Digital Sky Survey of $\sim$200,000
F and G stars (Ivezi\'c \etal\ 2008).  A circular velocity lag in halo
rotation has also been seen in external spiral galaxies (Collins,
Benjamin, \& Rand 2002).



\section{Data Analysis}
\subsection{Reduction and Analysis of Data}

Once the data were normalized and placed on a common velocity scale, we 
stacked plots of the \HI, \SiII, \SiIII, and \SiIV\ lines to observe whether 
the absorption features in each sight line share a velocity range.  A variety 
of examples, involving both IVCs and HVCs, are shown in Figure~3.  Although
we detected \SiII\ and \SiIV\ in several lines, we selected one line with the 
cleanest data to derive the column density. 
We set the velocity ranges by hand in order to minimize the errors caused by 
weak or saturated absorption features.  
To define the velocity ranges, we typically found \SiIII\ $\lambda1206.50$ 
to be the most useful, since it was always present and often quite strong.  
We used \SiII\ and \SiIV\ to identify velocity components and define the 
minimum velocity range, since their 
features, while harder to detect, did not have outer wings seen in the 
saturated \SiIII\ line.  The 21 cm line provided useful velocity 
limits, especially when \SiII\ and \SiIV\ features were not prominent 
or did not agree.


\begin{figure*}
\epsscale{1}
\plotone{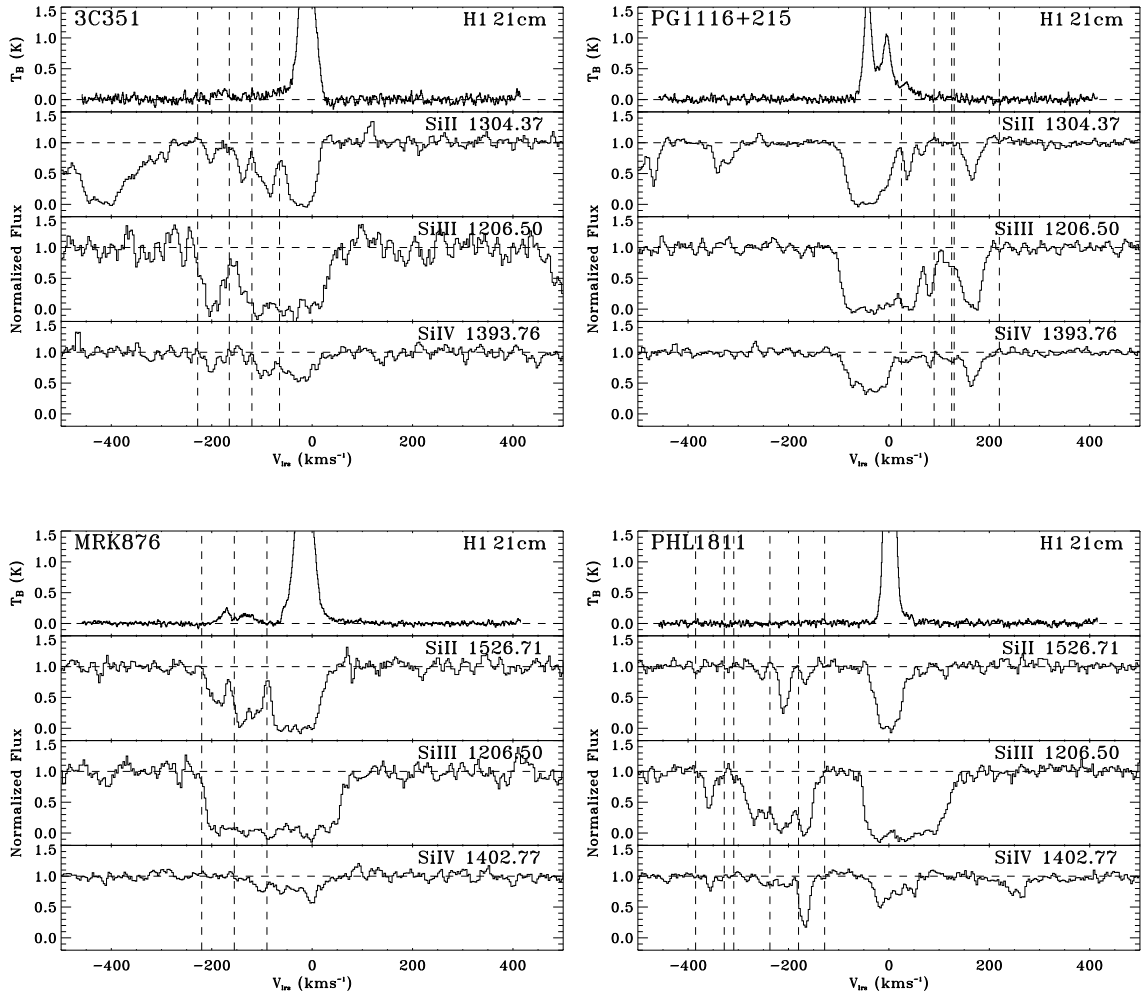}
  \caption{Normalized plots of HST/STIS spectra of 3C\,351, PG\,1116$+$215,
   Mrk\,876, and PHL\,1811, showing \SiII, \SiIII, and \SiIV\ absorption
   features and normalized plot of \HI\ emission feature from the LDS.  We mark
   with vertical dashed lines the LSR velocity range used to calculate the
   column densities. }
\end{figure*}


All 37 sight lines exhibited either an IVC or HVC absorption feature.  In 
7 sight lines, we detected no HVCs: 3C\,249, NGC\,4051, NGC\,4151, 3C\,273, 
NGC\,5548, Mrk\,1383, and IR\,2121$-$1757.  These statistics suggest 
an HVC covering factor of $81\pm5$\% (30 out of 37 sight lines) of the 
high-latitude Galactic sky in \SiIII, with an error bar based on an 
uncertainty of $\pm2$ sight lines in which we searched for \SiIII.
This \SiIII\ covering fraction is higher than the $\sim60$\%  seen in \OVI\ 
(Sembach \etal\ 2003), as expected from the relative strengths of \SiIII/\OVI.
Similar covering fractions of 80--90\% were found in our larger survey of 
\SiIII\ (Collins, Shull, \& Giroux 2009), in which 53 of 58 sight lines 
exhibited HVCs in \SiIII. 
In this paper, 8 of 37 sight lines exhibited three or more HVC/IVC components:  
Mrk\,335, NGC\,3783, PG\,0953$+$415, 3C\,351, Mrk\,509, PHL\,1811, UGC\,12163, and 
NGC\,7469.  In these and other multiple-component cases, we treated the HVCs 
and IVCs as individual absorbers.  However, in compiling statistics on the 
integrated column density of \SiIII, we combined the multiple HVC components 
into a single value per sight line. 

For each absorption feature, we measured the equivalent width and the 
column density in order to compare the observational data to the theoretical 
models (Table~3). The column density, $N_{\rm AOD}$, for each absorption feature 
was found using the apparent optical depth (AOD) method described in Savage \& 
Sembach (1991). Our software takes into account two sources of error:  
continuum placement and velocity limits of integration.  First, we moved the 
continuum up and down from the nominal placement by an amount equal to the the 
standard deviation of pixel values in the line-free regions, a relative change 
of $(S/N)^{-1}$.  The AOD measurements using these upper and lower continua 
generate upper and lower bounds for $N_{\rm AOD}$.  Second, we moved the limits 
of integration inward and outward from their nominal positions by 15 km s$^{-1}$ 
and measured the column density for a second set of upper and lower bounds.  For 
isolated lines, this resulted in negligible changes in $N_{\rm AOD}$, but for 
blended lines the difference can be significant.  Since the two effects are 
uncorrelated, we subtract the nominal $N_{\rm AOD}$ from the bounds and add the 
two upper and two lower uncertainties in quadrature for a final uncertainty.   

The equivalent width, $W_{\lambda}$, is related to the column density, but 
it was most important for finding the cases where the the column density is 
an upper limit.  The error in average equivalent width error at $\pm1\sigma$
and the average integration range error at the contracted and expanded 
velocity range were added in quadrature.  In setting upper limits on 
column density, we adopted $4 \sigma$ as a conservative standard and
used the relation,  
\begin{equation} 
   W_{min}=\frac{4\;\lambda_{0}} {R \; (S/N)} \; .
\end{equation}
Here, the instrumental resolution is $R =  \lambda/\Delta\lambda$, which we take 
to be $R = 15,000$ (\FUSE) and $R = 42,000$ (STIS/E140M).  In the cases where 
$W_{\rm min}$ was less than the measured equivalent width, 
we assume that the column density has only a $4\sigma$ significance.

The \HI\ column density was measured by integrating the 21-cm spin temperature 
over the velocity range $v_1 \leq v \leq v_2$ (in \kms), using the formula, 
\begin{equation}  
  N_{\rm HI} = (1.823 \times 10^{18}\;{\rm cm}^{-2}) \; 
         \int_{v_1}^{v_2} T(v) \;dv  \; ,
\end{equation}
where the factor $1.823 \times 10^{18}$ converts from velocity-integrated 
temperature (in K~\kms) to \HI\ column density (in \cm).  
The errors in \NHI\ are typically smaller than those for the Si ion column 
densities and were calculated by the formula,  
\begin{equation}
  \sigma_{\rm NHI} = (1.823 \times 10^{18}~{\rm cm}^{-2}) 
    (v_2-v_1) \sqrt{\frac{\Sigma (T-T_{mean})^2}{n-1}} \; ,
\end{equation}  
where $T$(K) is the spectral noise temperature, $(v_2-v_1)$ is the velocity
range (\kms), and $n$ is the number of temperature elements over that 
velocity range.  We have not included uncertainty in \NHI\ from a mismatch to
the LDS beam-size. Such effects can be important in some HVCs such as 
Complex C, with known sub-substructure, and have been estimated at
$\pm 0.1$ dex in log\,N$_{\rm HI}$ (Wakker \etal\ 2003; Collins \etal\ 2007).

\subsection{Integrated Column Densities of Ionized Gas}

The high ions of heavy elements (\CIII, \CIV, \NV, \SiIII, \SiIV,
\OVI) can be used as proxies for the amount of ionized hydrogen in the
low halo.  For example, the high-velocity \OVI\ absorbers (Sembach
\etal\ 2003) were found to have substantial mean column density, 
$\langle \log N_{\rm OVI} \rangle = 13.95 \pm 0.34$. Scaling from a 
fiducial metallicity of $0.2 Z_{\odot}$, we can estimate the corresponding
ionized hydrogen column density, 
\begin{eqnarray}
   {\rm N}_{\rm HII} & \approx & \frac { {\rm N}_{\rm OVI} }
    { {\rm (O/H)}_{\odot} \, f_{\rm OVI} \; (Z/Z_{\odot}) }  \nonumber \\ 
    & \approx & (4.5 \times 10^{18}~{\rm cm}^{-2}) (Z/0.2 Z_{\odot})^{-1} \; .
\end{eqnarray}
Here, we adopt (O/H)$_{\odot} = 4.9 \times 10^{-4}$ as the solar oxygen
abundance, assume that $f_{\rm OVI} \approx 0.2$ is the \OVI\ ionization
fraction in hot gas at $\log T_{\rm max} = 5.45$ (Sutherland \& Dopita 1993), 
and define $Z/Z_{\odot}$ as the oxygen metallicity relative to
solar abundances.  This high-velocity gas is likely to have metallicity in
the range $Z_{\rm HVC} =$ 0.1--0.3~$Z_{\odot}$,
similar to that of the Magellanic Stream (Lu, Savage, \& Sembach 1994;
Gibson \etal\ 2000; Sembach \etal\ 2001) or Complex~C (Wakker \etal\ 1999;
Gibson \etal\ 2001; Richter \etal\ 2001; Tripp \etal\ 2003;
Collins, Shull, \& Giroux 2003, 2007).

A more recent survey (Fox \etal\ 2006) of \OVI\ and \CIII\ examined 47 
highly ionized HVCs (without 21-cm emission) at 
100 \kms\ $<\, |v_{\rm VSR}| \, <$ 400~\kms. This HVC population
had mean column density $\langle {\rm N}_{\rm OVI} \rangle = 13.83 \pm 0.36$
and mean line width $\langle b_{\rm OVI} \rangle = 38 \pm 10$~\kms.  In 81\% 
(30 of 37) highly ionized HVCs, they found clear accompanying \CIII\ absorption
with a ratio, N(\CIII)/N(\OVI), which they claim is 
consistent with relative solar abundances and
collisional ionization equilibrium at $T = 1.7 \times 10^5$~K.
However, owing to the strong radiative cooling at these temperatures, the plasma 
is likely to be out of ionization equilibrium, which would alter the expected ratios.

The high-latitude ionized layer probed by the HVCs appears to have a total 
hydrogen column density $\sim10^{19}$~\cm, integrated through both sides 
of the Galactic plane.  Depending on its spatial distribution and distance above 
the plane, this gas could provide a significant reservoir of low-metallicity 
material, some of which may cool and fall onto the Galactic disk.  The total 
mass depends on the HVC distances and geometry.
If the vertical column density, $N_{\rm HII}$, in these HVCs resides in a thin
layer, above and below the Galactic disk with radius
$R \approx (10~{\rm kpc})R_{10}$, the total ionized mass is
\begin{eqnarray}
   M_{\rm HII} & \approx & (2 \pi R^2)(1.32 m_H) N_{\rm HII}  \nonumber \\
   & \approx & (3 \times 10^7~M_{\odot})\, R_{10}^2 \, 
       \left( \frac {Z} {0.2 Z_{\odot}} \right)^{-1} \; .
\end{eqnarray}
An HVC population around the Milky Way extending above a more extended disk 
($R \approx 15$~kpc) or with metallicity $\sim0.1~Z_{\odot}$ could have mass 
exceeding $10^8~M_{\odot}$.  

>From the \SiIII\ measurements, we now make the same calculation for the total 
hydrogen column density. As shown in Table~2, we detected \SiIII\ in 
83 absorption systems:  61 HVCs along 30 sight lines and 22 IVCs along 20 
sight lines.  Because the \SiIII\ line is saturated in several cases, we 
list {\it lower} limits on the column density, $\log {\rm N}_{\rm SiIII}$, 
computed by the AOD method.   In order to compare \SiIII\ to \OVI, we combine 
all HVCs in each sight line into a single \SiIII\ column density.  Averaging 
the total N(\SiIII) along each sight line using measured values or lower limits, 
we find $\langle \log N_{\rm SiIII} \rangle = 13.42 \pm 0.21$ (for 30 sight lines 
with \SiIII\ HVCs) and $13.59 \pm 0.25$ (for 20 sightlines with 22 IVCs).  
We derive the total hydrogen mass by correcting for the (solar) silicon abundance, 
(Si/H)$_{\odot} = 3.24 \times 10^{-5}$, with a typical HVC metallicity 
$Z_{\rm Si}/Z_{\odot} \approx 0.2$ (values of 10--20\% solar are observed in 
Complex C). We adopt an ionization fraction $f_{\rm SiIII} \approx 0.7\pm0.2$ 
characteristic of multiphase conditions. We estimate that the HVC population 
corresponds to a vertical (one-sided) column density of ionized hydrogen,
\begin{eqnarray}
   {\rm N}_{\rm HII} & \approx &\frac  { {\rm N}_{\rm SiIII} }
    { {\rm (Si/H)}_{\odot} \, f_{\rm SiIII} \; (Z_{\rm Si}/Z_{\odot})}  \nonumber \\  
   & \approx & (6 \times 10^{18}~{\rm cm}^{-2}) \left( \frac 
   {Z_{\rm Si}} {0.2 Z_{\odot}} \right)^{-1}  \left( \frac 
   {f_{\rm SiIII} } {0.7} \right)^{-1} \; .
\end{eqnarray}
Perhaps coincidentally, this column density is close to that
derived from \OVI\ (eq.\ [5]), with mean column density 
$\langle \log {\rm N}_{\rm OVI} \rangle = 13.95 \pm 0.34$ (Sembach \etal\ 2003), 
although these ions probably trace gas in different thermal phases.   

As noted by Collins \etal\ (2007), the hot gas at 10--20\% metallicity 
can cool on Gyr timescales, and the low-metallicity \OVI\
and \OVII\ reservoir might produce a mass inflow rate of at least 
$0.1~M_{\odot}$~yr$^{-1}$.  The \SiIII\ HVC absorbers observed in 
this survey probably have temperatures of $10^{4.0}$~K to $10^{4.5}$~K.  
A rough estimate of the mass infall rate comes by dividing 
the mass of ionized gas (scaled from \SiIII\ HVCs) by the time for HVCs
to fall from 10 kpc to the plane at 100--300 \kms.  This estimate can 
also be corrected for \HI\ using the typical ratio, 1.5--2.5, of total 
(\HI\ + \HII) to ionized hydrogen (Collins \etal\ 2003) and by a factor of 
0.5 for the fraction of HVCs falling inward.  Averaged over the Galactic disk 
of radius $R = 10-15$ kpc, the population of \SiIII\ HVCs has a total hydrogen 
mass $\sim 10^8~M_{\odot}$. Dividing by typical infall times of 30--100 Myr, 
we estimate an average mass infall rate $\sim 1~M_{\odot}~{\rm yr}^{-1}$.  
Although uncertain, this is a substantial fraction of the replenishment rate
needed to balance gas consumption by star formation in the disk (Gilmore 2001).

\subsection{Ionization Modeling}

Both HVCs and IVCs likely include multiphase gas, with a range of
ionization mechanisms.  Here, we examine the case of pure photoionization.  
For a subset of absorbers with the best data, including \SiII, \SiIII,
and \SiIV, totaling 17 HVCs and 19 IVCs, we were able to model the 
ionization conditions responsible for the silicon ions. 
In order to constrain the values of ionization parameter and metallicity,
we compared the observations to a grid of models calculated using 
{\it Cloudy} v96.02 last described by Ferland \etal\ (1998).  We assumed a 
plane-parallel geometry, with no dust, trace elements, or molecular hydrogen. 
We set the lifetime of the HVC/IVC equal to $10^8\rm{\,yr}$, much longer than 
equilibrium timescales, and we ignored the effects of cosmic rays.

\begin{figure*}[ht] 
  \includegraphics[angle=90,scale=.75]{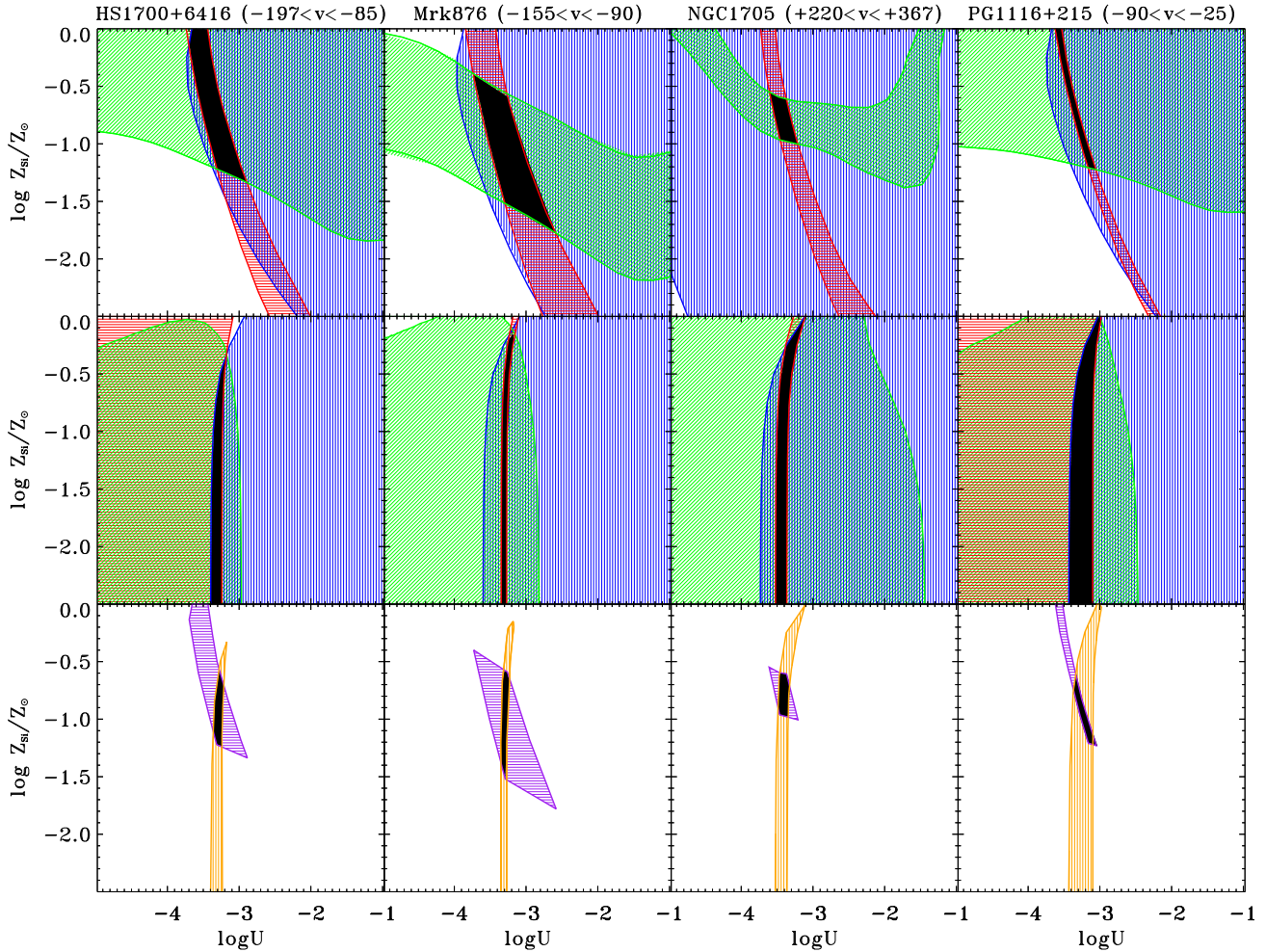}
  \caption{Photoionization models of the column densities of three Si ions constrain the ionization parameter, $\log\,U$, and silicon metallicity, $Z_{\rm Si}$, for three HVCs (toward HS\,1700$+$6417, Mrk\,876, NGC\,1705) and one IVC toward PG\,1116$+$215. The color scheme in these panels shows regions constrained by the individual Si ion ratios and illustrate where model ion ratios match ($1\sigma$) observed ratios.  In the top plots, we use log(\NSiII/\NHI), log\,(\NSiIII/\NHI), and log\,(\NSiIV/\NHI).  In middle plots, we show log\,(\NSiIII/\NSiIV), log\,(\NSiIII/\NSiII), and log\,(\NSiII/\NSiIV).  The bottom plots compare solutions from the top and middle plots.  Black regions show overlapping solutions. }
\end{figure*}

\begin{figure*}[ht] 
  \epsscale{.95}\plotone{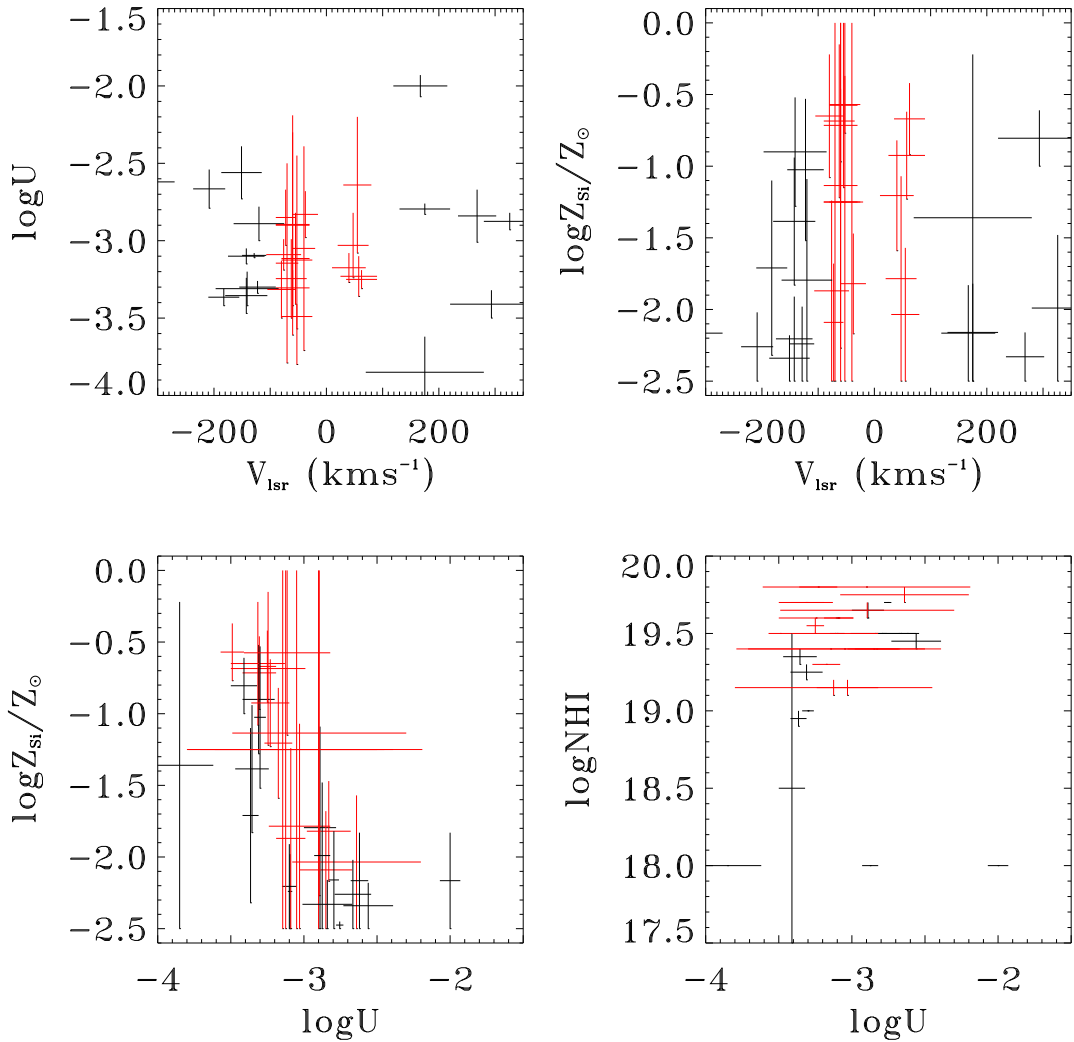}
  \caption{Correlation between different model parameters ($U$, $Z_{\rm Si}$) and absorber LSR velocity ($v_{\rm LSR}$) and column density (log\,\NHI).  Red points show models for IVCs, and black are for HVCs. We see that log\,$U$ is fairly constant, with little dependence on \NHI, $Z_{\rm Si}$, or $v_{\rm LSR}$. We see an apparent tilt in the relation between $Z_{\rm Si}$ and log $U$, which may arise from higher densities in IVCs compared to HVCs. }
\end{figure*}

\begin{figure*}
  \epsscale{.9}\plottwo{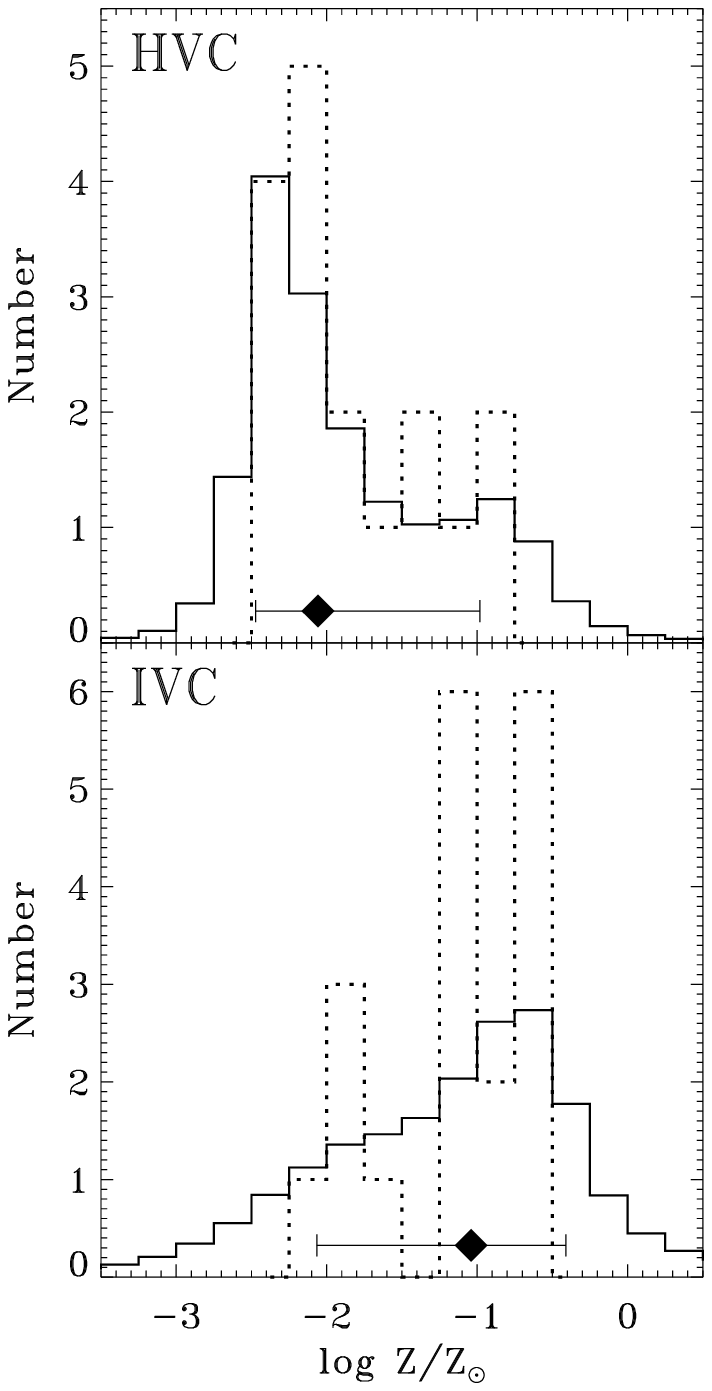}{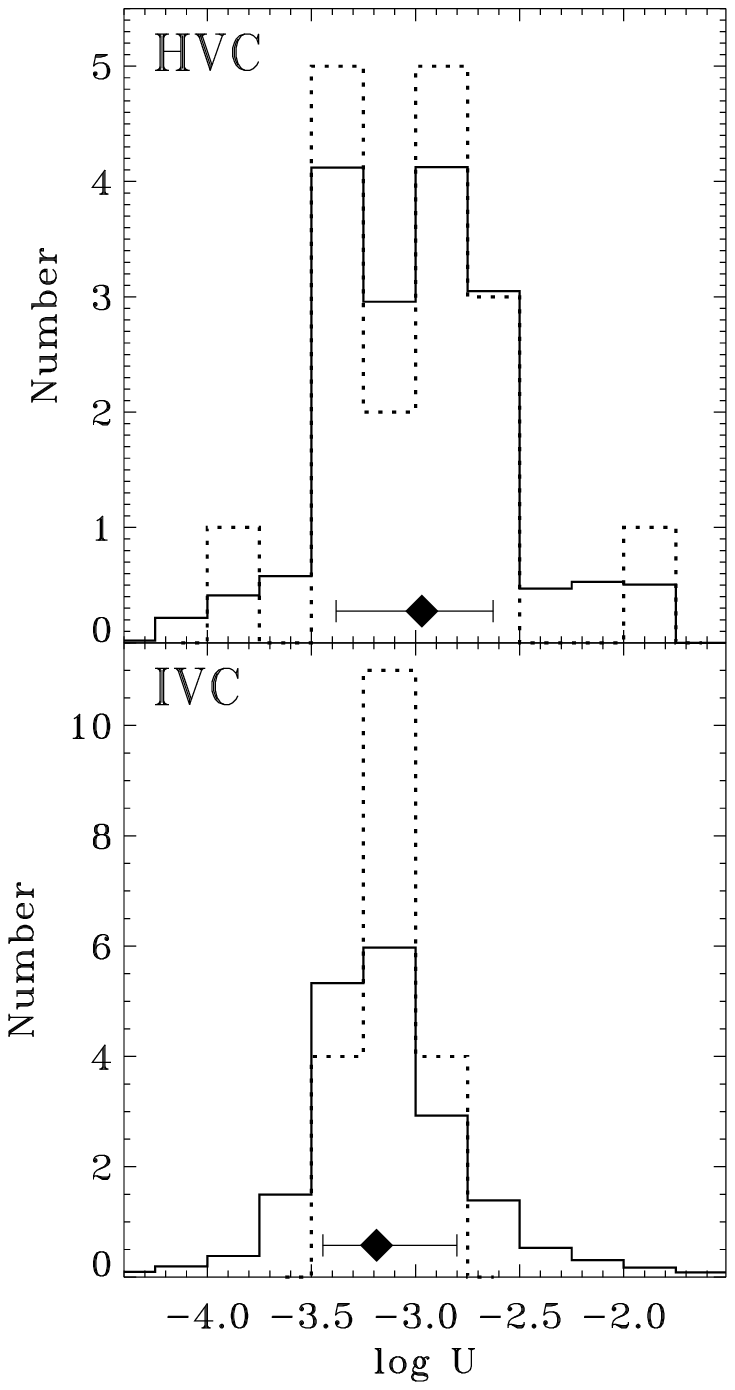}
  \caption{Histograms of $\log\,U$ and $\log\,(Z_{\rm Si}/Z_\sun)$ for 17 modeled HVCs 
  (top panels) and 19 modeled IVCs (bottom panels), assuming pure photoionization. 
  More reliable metallicities of 10--30\% solar can be derived from [\SiII/\HI].  
  Our multi-bin histogram technique 
  (see text) accounts for the distribution of metallicities and ionization parameters 
  into adjacent bins, assuming a Gaussian distribution.  Diamonds show medians (with 
  $1 \sigma$ variance in cumulative distribution).  The median ionization parameter is 
  well determined at $\log\,U = -3.0^{+0.3}_{-0.4}$ for HVCs, while metallicities are
  highly uncertain, with a broad range from $\log\, Z_{\rm Si}/Z_{\odot} = -2.5$ to $0$, 
  with median $\langle \log (Z_{\rm Si}/Z_{\odot}) \rangle = -2.06^{+1.08}_{-0.41}$ for 
  17 modeled HVC, and $-1.04^{+0.63}_{-1.03}$ for 19 modeled IVCs.}
\end{figure*}

The most important parameter in our photoionization models is the
photoionization parameter, $U = n_{\gamma}/n_H$, where $n_{\gamma}$
is the number density of ionizing photons and $n_H$ is the number density
of hydrogen in all forms.  For both AGN and OB-association sources, we
assumed an ionizing radiation field parameterized by specific intensity
$J_{\nu} = J_0 (\nu/\nu_0)^{-\alpha}$ in units 
erg~cm$^{-2}$ s$^{-1}$ Hz$^{-1}$ sr$^{-1}$.  For metagalactic radiation, 
the specific intensity at 1 ryd (Shull \etal\ 1999) 
lies in the range $J_0 \approx (1-2) \times 10^{-23}$, which 
corresponds to a normally incident (one-sided) photon flux of
$\Phi_0 \approx 10^4$ cm$^{-2}$~s$^{-1}$.  The mean spectral index 
is chosen as $\langle \alpha \rangle = 1.5$, consistent with the  
observed spectral indices (rest-frame 1--3 ryd) for AGN at 
redshifts $z \geq 0.3$ (Telfer \etal\ 2002) modified by transmission and 
reprocessing through the IGM (Fardal \etal\ 1998).  The shape of metagalactic 
ionizing spectrum is still uncertain, as indicated by the fact that composite 
spectra of AGN at $z < 0.3$ observed by \FUSE\ appear to have somewhat 
harder spectra with $\langle \alpha \rangle = 0.5-0.6$ 
(Scott \etal\ 2004; Shull \etal\ 2009). 

For HVCs within the Galactic halo, the radiation field is expected to 
be incident primarily from one side, dominated by OB-star radiation 
escaping from the Galactic disk.  This Galactic ionizing radiation field
is estimated to be 10-100 times stronger than the metagalactic field 
(Giroux \& Shull 1997).  We scale the normally incident flux of 
hydrogen-ionizing photons to the value, 
$\Phi_0 = (10^{5.5}$ \,photons \cm\,s$^{-1}) \Phi_{5.5}$, typical of gas 
in the low Galactic halo (Dove \& Shull 1994; Bland-Hawthorne \& Maloney 1999). 
Integrating over frequency and angle, we can relate $J_0$ to the number 
density, $n_{\gamma}$, of ionizing photons, the one-sided photon flux $\Phi_0$, 
and the photoionization parameter, $U = n_{\gamma}/n_H$:
\begin{eqnarray}
  \Phi_0  &=& \left( \frac {2 \pi J_0} {h c \, \alpha} \right) = 
           (10^{5.5}~{\rm photons~cm}^{-2}\,{\rm s}^{-1}) \, \Phi_{5.5} \; ,   \\
  n_{\gamma} &=& \frac {\Phi_0}{c} = (1.05 \times 10^{-5}~{\rm cm}^{-3}) \; 
                \Phi_{5.5} \, \left( \frac {1.5} {\alpha} \right)  \; , \\
    U       &=& \frac {n_{\gamma}} {n_H} = (1.05 \times 10^{-3}) \, \Phi_{5.5}  
                   \, n_{-2}^{-1} \left( \frac {1.5} {\alpha} \right) \; . 
\end{eqnarray} 
Here, we scaled the HVC hydrogen density to typical values,
$n_H = (10^{-2}~{\rm cm}^{-3}) n_{-2}$.  For instance, our photoionization 
models of multiple ion stages in Complex~C (Gibson \etal\ 2001; Collins \etal\ 
2003, 2007) find $n_H \approx$ 0.01--0.1 cm$^{-3}$.  

For a normally incident radiation field, the photon flux is
$\Phi_0 = n_{\gamma} c$, while an isotropic radiation field gives 
$\Phi_0 = n_{\gamma} c/4$.  Thus, to relate photoionization parameter 
$U$ to $n_{\gamma}$, one must specify the geometry of the radiation field
(one-sided normally-incident or diffuse isotropic).  We explored a range 
$-4 < \log U < -1$ and constructed a three-dimensional grid of models 
with parameters: total hydrogen density $n_H$, neutral hydrogen column density 
\NHI, and silicon metallicity $Z_{\rm Si}/Z_\sun$.  
We varied the metallicity over the range $-2.5 < \log (Z_{\rm Si}/Z_\sun) < 0$ 
and the \HI\ column density over the range $17 < \log$ \NHI\,(cm$^{-2}$) $<20$ 
in logarithmic steps of 0.1 dex.  We varied the hydrogen density from 
$10^{-4} < n_H\, ({\rm cm}^{-3}) < 1$, spanning a broad range in density 
from diffuse intergalactic medium (IGM) to diffuse interstellar medium (ISM).  
Each model produced column densities for many species, including \SiII, \SiIII, 
\SiIV, which are studied relative to \HI.  We then compared 
model column density ratios to the observed ratios for all absorbers (HVCs and 
IVCs).  From the observed \NHI, we then found solutions for $n_H$ and 
$Z_{\rm Si}/Z_{\odot}$. The quoted error bars reflect the range in parameters 
($U$ and $Z_{\rm Si}$) allowed by the photoionization models.  We have not 
included systematic errors arising from assumptions about the spectral shape of 
the photoionizing radiation field, or the relative contributions of collisional
ionization from hot gas.  

The question of how to properly model the ionization corrections for multiphase 
HVCs is complicated and beyond the scope of this paper.  One possible approach 
to address this issue would use multiple ion states from another element 
(\CII, \CIII, \CIV) to constrain the ionization ratio (photoionized/collisional).  
Another approach is to use the ``neutral-gas'' ratios, [\OI/\HI], [\SII/\HI], 
[\SiII/\HI], to derive the metallicity, and then assume that this metallicity 
applies to the higher ions, \SiIII, \SiIV, \CIII, \CIV.  We will explore
these approaches in a subsequent paper.  For now, we simply note that the 
photoionization modeling done here introduces systematic uncertainties
in the HVC (silicon) metallicities, by the lack of any correction for
\SiIII\ and \SiIV\ produced in hot, collisionally ionized gas.   
Thus, the low values of $Z_{\rm Si}$ described in \S~4 may reflect the 
possibility that \SiIII\ and \SiIV\ are not co-spatial with the \HI.   
 


\section{Results and Discussion}

The column density ratios of the three silicon ions relative to \NHI\ 
were compared to photoionization models to match the HVC/IVC characteristics. 
The ionization parameter, $\log U$, is best constrained by the pairs
of individual silicon ions, \SiII, \SiIII, and \SiIV.  Once we have
a solution for $\log U$, we constrain the Si metallicity by the ratio
of silicon column densities to $\log N_{\rm HI}$.  The \SiIII\ did not 
provide as much of a constraint on the model ratios as the weaker \SiII\ 
and \SiIV\ ions.  However, because of its strength and sensitivity, \SiIII\ 
was used to identify HVCs and define their velocity ranges.
These models yield a range of silicon metallicity ($Z_{\rm Si}$) and 
hydrogen density ($n_H$).   Because $U \propto \Phi_H/n_H$, we can then 
infer the photoionizing flux, $\Phi_H$.  From $U$ and $Z_{\rm Si}$, we can 
estimate the location of the HVCs.  Sample solutions for $U$ and $Z_{\rm Si}$ in 
four absorbers (three HVCs and one IVC) are displayed in Figure~4,
which illustrates our method for determining $\log U$ and $\log Z_{\rm Si}$. 

The full solutions are given in Table~3, and Figure~5 displays the
correlations among metallicity, ionization parameter, velocity, and
\NHI.  We observe an apparent tilt in the relation between $Z_{\rm
Si}$ and log $U$, which may arise from higher densities in IVCs
compared to HVCs.  Figure~6 presents a set of histograms that show the
range of ionization parameter $U$ and metallicity $Z_{\rm
Si}/Z_{\odot}$.  Because we can determine $U$ more accurately than
$Z_{\rm Si}$, we have plotted these quantities using a multi-bin
histogram technique that distributes power into adjacent bins assuming
a Gaussian distribution; see Danforth \& Shull (2008) for more
details.

The models for Si metallicity spanned the range $-2.5 \leq \log
(Z_{\rm Si}/Z_\sun) \leq 0$, with mean values $\langle \log (Z_{\rm
Si}/Z_{\odot}) \rangle = -2.1^{+1.1}_{-0.4}$ (for 17 modeled HVCs) and
$-1.0^{+0.6}_{-1.0}$ (for 19 modeled IVCs).  These metallicities are
somewhat lower than the mean (column-density weighted) value, $\langle
\log (Z_{\rm O}/Z_{\odot}) \rangle = -0.89$ found for Complex~C
(Collins \etal\ 2007).  They are also lower than the metallicities of
``neutral gas'' estimated from [\SiII/\HI] and of the LMC and SMC.
However, they agree fairly well with the mean metallicities, [Fe/H] =
$-1.46 \pm 0.30$ measured for $\sim$200,000 halo F/G stars in the
Sloan Digital Sky Survey (Ivezi\'c \etal\ 2008).

The range of metallicities in IVCs extends to higher values than in
the HVCs.  In the lowest-column bins of IVCs, three of the four
absorbers could be a part of a larger HVC system.  This distinction
may not be significant, owing to the definition used to separate HVCs
and IVCs at 90--100\,\kms.  Within the set of models, the ionization
parameter ranged from $-4 \leq \log\, U \leq -1$.  The solutions for
the 61 HVCs and 22 IVCs showed that most could be explained by $\log U
\approx -3.0 \pm 0.3$.  Similar modeling for absorbers in the
low-redshift IGM (Danforth \& Shull 2005, 2008) found $\log U \approx
-2.1 \pm 0.5$, for a diffuse, isotropic radiation field.  Thus, the
HVCs appear to have an ionization parameter ten times lower than that
of the IGM, suggesting that HVCs reside in the higher-pressure
Galactic halo, with a larger $n_H$.

Whether the HVC metallicity differences between the silicon-ion
photoionization models and [\OI/\HI] are statistically significant may
depend on systematic uncertainties in our photoionization modeling.
These models are based on ratios of \SiII, \SiIII, and \SiIV\ column
densities (from UV absorption lines) to \NHI\ determined from 21-cm
emission.  Beam-size effects could make a difference in matching the
\HI\ to the absorbing sight line.  We believe the mean ionization
parameter to be fairly accurately determined at $\log U \approx -3.0
\pm 0.3$, but the range of metallicities is somewhat larger. The
typical uncertainties for individual sight lines are $\pm 0.1$ in
$\log U$ and $\pm 0.5$ in metallicity $\log Z_{\rm Si}$.


\medskip 
\noindent
We now summarize the main results of our survey. 
\begin{enumerate}

\item The strong \SiIII\ 1206.50 \AA\ absorption line is a sensitive tracer
    of high-velocity and intermediate-velocity ionized gas, typically
    with 4--5 times higher optical depth than \OVI\ $\lambda 1031.93$.
    With high-resolution UV spectrographs, this allows sensitivity to
    HVCs with \NSiIII\ as low as $5 \times 10^{11}$ \cm, corresponding
    to ionized hydrogen column densities $N_H \approx 8 \times
    10^{16}$ \cm\ at $0.2 Z_{\odot}$ metallicity.

\item The HVC covering factor of the high-latitude Galactic sky in \SiIII\
    is $81\pm5$\%, somewhat higher than seen in in \OVI\ (60\%) but
    much larger than in \HI\ 21-cm emission (37\%). The mean HVC
    column density per sight line is $\langle \log N_{\rm SiIII}
    \rangle = 13.42 \pm 0.21$ for 30 sight lines with 61 HVCs and
    $13.59 \pm 0.25$ for 20 sight lines with 22 IVCs.
 
\item Correcting for silicon metallicity and ionization fraction, we
    estimate that the \SiIII\ corresponds to a column density of
    ionized hydrogen, N$_{\rm HII} \approx (6 \times 10^{18}~{\rm
    cm}^{-2}) (Z_{\rm Si}/0.2Z_{\odot})^{-1}$, similar to that
    inferred from the population of \OVI\ HVCs (Sembach \etal\ 2003).
    If this ionized HVC layer extends $\sim10$~kpc above and below the
    Galactic plane, it could contain $10^8~M_{\odot}$ of
    low-metallicity (10--20\% solar) gas. The infall times for HVCs at
    10 kpc and 100--200 \kms\ are 50--100 Myr. This infall averages
    $1~M_{\odot}$~yr$^{-1}$, a substantial portion of the replenishment 
    rate for star formation in the disk.

\item Using all three silicon ions (\SiII, \SiIII, \SiIV) and assuming
    pure photoionization, we infer a mean photoionization parameter, 
    $\log U \approx -3.0 \pm 0.2$, and average (highly uncertain) silicon 
    metallicity, $\langle \log (Z_{\rm Si}/Z_\sun) \rangle = -1.4 \pm 0.3$
    ($-1.8 \pm 0.4$ for HVCs and $-1.2 \pm 0.4$ for IVCs). This ionizing 
    radiation field is $\sim10$ times lower than inferred in the low-$z$ 
    IGM.  The HVC metallicities inferred from [\SiII/\HI] are 10--30\%. 
    The lower metallicities found from all three ions are unreliable, 
    because of the likely contributions from collisional ionization.   

\item We see HVC velocity segregation on either side of the rotation axis,
    with blue-shifted absorbers ($V_{\rm LSR} < -90$~\kms) found primarily at 
    $\ell < 180^{\circ}$ and red-shifted gas ($V_{\rm LSR} > +90$~\kms) at
    $\ell > 180^{\circ}$. This effect has been seen in other species 
    (Collins \etal\ 2005; Fox \etal\ 2006) and may result from a lag in the
    circular rotation velocity at elevations off the disk plane. 

\end{enumerate} 

\noindent In the future, we eagerly await the installation of the {\it Cosmic
Origins Spectrograph} in Servicing Mission 4 for the {\it Hubble Space
Telescope}.  This mission may also repair STIS.  Together, these UV
spectrographs will rejuvenate spectroscopic studies of HVCs and IVCs
towards many AGN targets.  With COS, we have targeted ten sight lines
toward HVCs in Complex~C, Complex A, WD, WB, and several
negative-velocity \OVI\ HVCs.  These studies, at S/N $\approx$ 30--40
and 15 \kms\ resolution, will survey physical conditions and
abundances in cool, warm, and hot HVC gas.  Based on our experience in
this paper, we believe that the silicon ions will play an important
role in defining the ionization conditions and HVC/IVC kinematics.

\acknowledgments

Our group's research support at the University of Colorado for UV
studies of the IGM and Galactic halo gas comes from COS grant
NNX08-AC14G and STScI spectroscopic archive grants (AR-10645.02-A and
AR-11773.01-A).  We also have support from NSF grant AST07-07474.  We
thank Steve Penton and Jason Tumlinson for reducing the initial STIS
and \FUSE\ data and Gary Ferland and Mark Giroux for useful
discussions regarding the CLOUDY modeling. We also thank the referee
for insights and comments that improved our arguments.  This work
contains data obtained for the Guaranteed Time Team by the
NASA-CNES-CSA {\it FUSE} mission operated by the Johns Hopkins
University, as well as data from the {\it Hubble Space Telescope}.


\clearpage

\clearpage



\LongTables
\begin{deluxetable}{lcccccc}
\tabletypesize{\scriptsize}
\tablecolumns{7}
\tablewidth{0pt}
\tablecaption{HVC and IVC Sight-line\tablenotemark{a} Statistics}
\tablehead{\colhead{Sightline}        &
         \colhead{Velocity Range}   &
         \colhead{Absorber}         &
         \colhead{log\,{\NHI}}      &
         \colhead{log\,{\NSiII}}    &
         \colhead{log\,{\NSiIII}}   &
         \colhead{log\,{\NSiIV}}    \\
         \colhead{}                 &
         \colhead{(\kms)}           &
         \colhead{Type}             &
         \colhead{(N in \cm)}       &
         \colhead{(N in \cm)}       &
         \colhead{(N in \cm)}       &
         \colhead{(N in \cm)}      }

\startdata
Mrk\,335       & $-$431, $-$385 & HVC & $<$19.37                & 
      12.71$^{+0.15}_{-0.14}$ & 12.41$^{+0.11}_{-0.06}$ & $<$12.25\\
             & $-$365, $-$280 & HVC & $<$19.37                & 
      14.24$^{+0.12}_{-0.28}$ & 12.99$^{+0.05}_{-0.03}$ & $<$12.47\\
             & $-$264, $-$238 & HVC & $<$19.37                & 
      $<$12.58                & 12.29$^{+0.10}_{-0.10}$ & 
      12.47$^{+0.24}_{-0.12}$\\
             & $-$137, $-$85  & HVC & $<$19.37                & 
      12.96$^{+0.11}_{-0.13}$ & $\geq$13.44             & $<$12.25\\
             & $-$85, $-$25   & IVC & 19.48$^{+0.10}_{-0.13}$ & 
      14.74$^{+0.15}_{-0.12}$ & $\geq$13.63   & 12.97$^{+0.08}_{-0.10}$\\

Ton\,S210      & $-$251, $-$198 & HVC & $<$19.66                & 
     $<$12.57  & 12.88$^{+0.08}_{-0.09}$ & 12.80$^{+0.08}_{-0.05}$\\
             & $-$198, $-$127 & HVC & $<$19.66  & 13.19$^{+0.08}_{-0.01}$ 
      & $\geq$13.68             & $<$12.47\\

HE\,0226$-$440 & 50, 90         & IVC & $<$18.00                & $<$12.59                
      & 12.10$^{+0.11}_{-0.30}$ & $<$12.05\\
      & 119, 215       & HVC & $<$18.00                & 12.35                   
      & 13.18$^{+0.09}_{+0.05}$ & 12.71$\pm0.14$\\

PKS\,0312$-$770& 70, 280        & HVC & $<$18.00                & $\geq$14.93             
        & $\geq$14.32             & 13.28$^{+0.07}_{-0.05}$\\
        & 280, 372       & HVC & $<$18.00                & 13.29$\pm0.06$          
        & $\geq$13.61             & 13.28$\pm0.06$\\

PKS\,0405$-$12 & 110, 170       & HVC & $<$19.52                & $<$12.55                
        & 12.50$\pm0.09$          & 12.64$^{+0.16}_{-0.16}$\\

NGC\,1705      & 90, 143        & HVC & $\geq$19.67             & $\geq$14.63             
     & 12.69$^{+0.13}_{-0.09}$ & $<$12.07\\
           & 210, 357       & HVC & 17.48$\pm0.20$          & 13.55$\pm0.04$          
     & $\geq$13.67             & 12.80$^{+0.16}_{-0.14}$\\

HS\,0624$+$6907 & $-$150, $-$90  & HVC & 19.19$^{+0.28}_{-1.00}$ & 14.15$^{+0.14}_{-0.12}$ 
     & 13.13$^{+0.12}_{-0.20}$ & $<$12.54 \\
     & $-$90, $-$60 & IVC & $<$19.63 &  $\geq$13.91 & $\geq$13.41 
        & 12.71$^{+0.31}_{-0.25}$ \\ 
     & 100, 135  & HVC & $<$19.63 & $<$12.43 & 12.26$^{+0.23}_{-0.12}$ & $<$12.00 \\

PG\,0953$+$415 & $-$185, $-$145 & HVC & $<$19.32                & 12.76$^{+0.46}_{-0.13}$ 
& 12.41$^{+0.13}_{-0.16}$ & 12.71$^{+0.24}_{-0.15}$\\
             & $-$145, $-$100 & HVC & $<$19.32                & $<$12.30                
     & $\geq$13.31      & 12.66$^{+0.22}_{-0.14}$\\
             & 95, 150        & HVC & $<$19.32                & $<$12.30    
             & 12.73$^{+0.08}_{-0.05}  $ & $<$12.05\\
             & 150, 185       & HVC & $<$19.32                & $<$12.30       
             & 11.92$^{+0.33}_{-0.28}  $ & $<$12.40\\

Ton\,28        & 95, 180        & HVC & $<$19.53                & $<$12.70    
             & 12.75$\pm0.07$ & 12.92$^{+0.13}_{-0.12}$\\

3C\,249        & $-$80, $-$30   & IVC & 19.72$^{+0.07}_{-0.08}$ & $\geq$14.61    
             & $\geq$13.92      & 13.29$^{+0.05}_{-0.09}$\\

NGC\,3516      & $-$197, $-$125 & HVC & $<$19.44                & 
     12.98$^{+0.17}_{-0.15}$ & $\geq$13.26  & 12.86$^{+0.13}_{-0.19}$\\
             & $-$90, $-$30   & IVC & 19.83$\pm0.06$          & $\geq$14.59    
             & $\geq$13.74      & 13.28$^{+0.11}_{-0.07}$\\

PG\,1116$+$215 & $-90$, $-25$   & IVC & 19.83$^{+0.06}_{-0.07}$ & $\geq$14.44       
       & $\geq$13.80      & 13.24$^{+0.06}_{-0.08}$\\
       & 25, 90         & IVC & 19.13$^{+0.25}_{-0.66}$ & 13.42$^{+0.04}_{-0.09}$ 
       & $\geq$13.61      & 12.59$^{+0.09}_{-0.15}$\\
       & 90, 130        & HVC & $<$19.49                & $<$12.22      
       & 12.70$^{+0.11}_{-0.04}$ & 12.39$^{+0.19}_{-0.09}$\\
       & 130, 220       & HVC & $<$19.49                & 13.58$^{+0.02}_{-0.02}$ 
       & $\geq$13.44      & 13.10$^{+0.03}_{-0.03}$\\

NGC\,3783      & 105, 158       & HVC & $\geq$20.00             & 
     13.14$^{+0.11}_{-0.06}$ & 12.95$^{+0.05}_{-0.12}$ & 12.20$^{+0.12}_{-0.06}$\\
             & 158, 208       & HVC & 19.18$\pm0.70$          & 
     13.05$^{+0.03}_{-0.05}$ & 12.58$^{+0.04}_{-0.03}$ & $<$11.59\\
     & 208, 293       & HVC & 20.07$\pm0.90$          & 14.27$^{+0.02}_{-0.02}$ 
     & 12.86$^{+0.04}_{-0.02}$ & 12.64$^{+0.09}_{-0.01}$\\
NGC\,4051      & $-$90, $-$50   & IVC & 19.34$^{+0.11}_{-0.14}$ & $\geq$14.45             & $\geq$13.76      
       & 13.20$^{+0.01}_{-0.11}$\\
NGC\,4151      & $-$90, $-$50   & IVC & 19.42$^{+0.11}_{-0.16}$ & $\geq$14.67             & $\geq$13.57      
       & 12.90$^{+0.16}_{-0.19}$\\
PG\,1211$+$14  & 35, 90         & IVC & $<$19.49                & 13.41$^{+0.13}_{-0.08}$ & 12.87$^{+0.18}_{-
0.08}$ & 12.07$^{+0.19}_{-0.16}$\\
             & 135, 185       & HVC & $<$19.49                & 12.81$^{+0.12}_{-0.18}$ & 12.36$^{+0.12}_{-
0.05}$ & $<$11.76\\
             & 240, 275       & HVC & $<$19.49                & $<$12.13                & 12.09$^{+0.14}_{-
0.10}$ & $<$11.76\\
PG\,1216$+$09  & 148, 219       & HVC & $<$19.48                & $<$12.73                & 12.87$^{+0.09}_{-
0.07}$ & 12.99$^{+0.14}_{-0.11}$\\
             & 219, 287       & HVC & $<$19.48                & $<$12.73                & $\geq$13.20      
       & $<$12.35\\
Mrk\,205       & $-$225, $-$175 & HVC & 19.44$^{+0.13}_{-0.19}$ & 12.87$^{+0.11}_{-0.14}$ & 12.43$^{+0.16}_{-
0.17}$ & $<$12.11\\
             & $-$175, $-$110 & HVC & 18.11$^{+0.09}_{-0.11}$ & 13.35$^{+0.06}_{-0.05}$ & 13.22$^{+0.10}_{-
0.07}$ & 12.47$\pm0.15$\\
3C\,273        & 10, 70         & IVC & 19.31$^{+0.19}_{-0.34}$ & 14.15$^{+0.11}_{-0.18}$ & $\geq$13.77      
       & 13.13$^{+0.07}_{-0.14}$\\
Q\,1230$+$011  & 80, 141        & HVC & $<$19.67                & 13.00$^{+0.18}_{-0.10}$ & 12.61$\pm0.07$   
       & $<$11.93\\
             & 274, 307       & HVC & $<$19.67                & 13.41$^{+0.08}_{-0.11}$ & 12.54$\pm0.07$   
       & $<$12.33\\
NGC\,4593      & 20, 75         & IVC & 19.14$^{+0.25}_{-0.68}$ & 13.83$^{+0.11}_{-0.28}$ & $\geq$13.54      
       & 13.01$^{+0.13}_{-0.21}$\\
             & 82, 112        & HVC & $<$19.52                & 12.83$^{+0.13}_{-0.10}$ & 12.43$^{+0.12}_{-
0.08}$ & $<$12.23\\
             & 252, 285       & HVC & $<$19.52                & 13.06$\pm0.10$          & 12.29$^{+0.10}_{-
0.01}$ & $<$12.50\\
PG\,1259$+$59  & $-$155, $-$95  & HVC & 19.37$^{+0.21}_{-0.42}$ & 14.05$^{+0.14}_{-0.13}$ & $\geq$13.60      
       & 12.67$^{+0.06}_{-0.07}$\\
             & $-$80, $-$30   & IVC & 19.51$^{+0.16}_{-0.27}$ & $\geq$14.77             & $\geq$13.83      
       & 12.92$\pm0.06$\\
PKS\,1302$-$102& 200, 340       & HVC & $<$19.56                & $<$12.48                & 12.23$^{+0.17}_{-
0.15}$ & 12.80$^{+0.19}_{-0.16}$\\
Mrk\,279       & $-$226, $-$110 & HVC & 19.28$^{+0.23}_{-0.50}$ & 13.94$^{+0.08}_{-0.10}$ & $\geq$14.05      
       & 12.57$^{+0.24}_{-0.22}$\\
             & $-$100, $-$55  & IVC & 19.66$^{+0.11}_{-0.15}$ & $\geq$14.54             & $\geq$13.65      
       & 13.29$^{+0.10}_{-0.14}$ \\
NGC\,5548      & $-$90, $-$25   & IVC & 19.19$^{+0.25}_{-0.69}$ & $\geq$14.84             & $\geq$13.74      
       & 13.68$\pm0.05$\\
Mrk\,1383      & $-$90, $-$55   & IVC & $<$19.43                & 13.53$^{+0.15}_{-0.20}$ & $\geq$13.50      
       & 13.00$^{+0.20}_{-0.11}$\\
PG\,1444$+$407 & $-$184, $-$126 & HVC & $<$19.50                & $<$12.60                & 12.62$^{+0.08}_{-
0.07}$ & $<$12.21\\
             & $-$107, $-$77  & IVC & $<$19.50                & 13.43$^{+0.10}_{-0.08}$ & $\geq$13.55      
       & 12.66$^{+0.18}_{-0.22}$\\
             & $-$77, $-$37   & IVC & 19.43$^{+0.14}_{-0.22}$ & $\geq$14.52             & $\geq$13.68      
       & 13.29$^{+0.16}_{-0.06}$\\
Mrk\,876       & $-$230, $-$155 & HVC & 18.91$^{+0.28}_{-1.1}$  & 13.67$^{+0.09}_{-0.07}$ & $\geq$13.61      
       & 12.56$^{+0.07}_{-0.08}$\\
             & $-$155, $-$90  & HVC & 18.97$^{+0.26}_{-0.71}$ & 14.27$^{+0.07}_{-0.04}$ & $\geq$13.92      
       & 13.20$^{+0.06}_{-0.03}$\\
HS\,1700$+$6416& $-$187, $-$85  & HVC & 19.23$^{+0.22}_{-0.45}$ & $\geq$14.36             & $\geq$14.03      
       & 13.13$^{+0.13}_{-0.12}$ \\
3C\,351        & $-$246, $-$183 & HVC & $<$19.61                & 13.11$\pm0.07$          & $\geq$13.50      
       & 12.69$^{+0.09}_{-0.08}$ \\
             & $-$183, $-$138 & HVC & $<$19.61                & 13.38$^{+0.11}_{-0.01}$ & 12.83$^{+0.22}_{-
0.12}$ & $<$12.07\\
             & $-$138, $-$83  & HVC & $<$19.61                & 13.87$^{+0.09}_{-0.08}$ & $\geq$13.72      
       & 12.94$^{+0.08}_{-0.06}$\\
H\,1821$+$643  & $-$188, $-$98  & HVC & $<$19.66                & 14.15$^{+0.05}_{-0.06}$ & $\geq$14.10      
       & 13.42$^{+0.03}_{-0.05}$\\
Mrk\,509       & $-$343, $-$261 & HVC & $<$19.66                & $<$12.58                & 13.30$\pm0.12$   
       & 13.27$\pm0.03$\\
             & $-$261, $-$218 & HVC & $<$19.66                & $<$12.58                & 12.47$^{+0.08}_{-
0.07}$ & 12.34$^{+0.10}_{-0.01}$\\
             & 35, 90         & IVC & 19.55$^{+0.16}_{-0.25}$ & 14.74$^{+0.07}_{-0.08}$ & $\geq$13.76      
       & 13.46$^{+0.04}_{-0.08}$\\
             & 100, 147       & HVC & $<$19.66                & 12.74$^{+0.12}_{-0.15}$ & 12.70$^{+0.09}_{-
0.12}$ & $<$12.47\\
IR\,2121$-$1757& 30, 80         & IVC & $<$19.76                & 13.86$^{+0.12}_{-0.11}$ & $\geq$13.84      
       & $\geq$13.68\\
PHL\,1811      & $-$390, $-$333 & HVC & $<$19.48                & $<$12.27                & 12.51$\pm0.04$   
       & 12.58$^{+0.07}_{-0.05}$\\
             & $-$314, $-$242 & HVC & $<$19.48                & $<$12.27                & 13.07$^{+0.07}_{-
0.03}$ & 12.77$^{+0.06}_{-0.07}$\\
             & $-$242, $-$190 & HVC & $<$19.48                & 13.39$\pm0.03$          & $\geq$13.38      
       & 12.95$\pm0.06$\\
             & $-$190, $-$133 & HVC & $<$19.48                & 12.69$^{+0.08}_{-0.09}$ & $\geq$13.43      
       & 13.51$^{+0.03}_{-0.02}$\\
PKS\,2155$-$304& $-$314, $-$220 & HVC & $<$19.47                & $<$12.17                & 12.33$^{+0.05}_{-
0.04}$ & 11.98$^{+0.15}_{-0.15}$\\
             & $-$200, $-$96  & HVC & $<$19.47                & 12.51$^{+0.16}_{-0.13}$ & 13.17$\pm0.02$   
       & 12.69$^{+0.04}_{-0.03}$\\
UGC\,12163    & $-$480, $-$390 & HVC & $<$19.69                & 13.64$^{+0.12}_{-0.14}$ & $\geq$13.91      
       & $<$12.54\\
             & $-$390, $-$330 & HVC & $<$19.69                & $<$12.83                & 12.95$\pm0.30$   
       & $<$12.84\\
             & $-$185, $-$135 & HVC & $<$19.69                & $<$13.09                & $\geq$13.16      
       & $<$12.84\\
             & $-$90, $-$35   & IVC & 19.60$^{+0.15}_{-0.23}$ & $\geq$14.47             & $\geq$13.75      
       & 13.36$^{+0.41}_{-0.27}$\\
NGC\,7469      & $-$419, $-$279 & HVC & $<$19.47                & 13.51$\pm0.08$          & $\geq$13.91      
       & 13.30$^{+0.07}_{-0.07}$\\
             & $-$279, $-$246 & HVC & $<$19.47                & $<$12.75                & $\geq$13.03      
       & 12.75$^{+0.15}_{-0.12}$\\
             & $-$199, $-$123 & HVC & $<$19.47                & $<$12.78                & 12.54$^{+0.15}_{-
0.12}$ & 12.87$^{+0.15}_{-0.12}$\\
             & $-$70, $-$20   & IVC & 19.21$^{+0.21}_{-0.41}$ & $\geq$14.65             & $\geq$13.38      
       & 13.16$^{+0.17}_{-0.08}$\\
\enddata
\tablenotetext{a}{Several velocity ranges are selected for many sight lines,
each treated as a separate HVC or IVC.}

\end{deluxetable}



\LongTables
\begin{deluxetable}{lccccc}
\tabletypesize{\scriptsize}
\tablecolumns{4}
\tablewidth{0pt}
\tablecaption{HVC and IVC Model Parameters\tablenotemark{a} }
\tablehead{\colhead{Sight line}   &
          \colhead{Velocity Range}  &
          \colhead{Absorber}     &
          \colhead{log\NHI}  &
          \colhead{$\log\, U$}        &
          \colhead{$\log\, (Z/Z_\sun)$} }
\startdata
Mrk\,335\tablenotemark{a}    & $-$426, $-$380 & HVC & 19.4       & \nodata          & \nodata \\
                           & $-$360, $-$275 & HVC & 19.4       & \nodata          & \nodata \\
                           & $-$257, $-$236 & HVC & 19.4       & \nodata          & \nodata \\
                           & $-$132, $-$80  & HVC & 19.4       & \nodata          & \nodata \\
                           & $-$80, $-$25   & IVC & 19.5       & $-$3.57, $-$3.41 & $-$0.77, $-$0.37 \\
Ton\,S210\tablenotemark{a}   & $-$266, $-$221 & HVC & 19.6, 19.7 & \nodata          & \nodata \\
                           & $-$221, $-$148 & HVC & 19.6, 19.7 & \nodata          & \nodata \\
HE0226$-$4410\tablenotemark{a}& 50, 90        & IVC & 18.0       & \nodata          & \nodata \\
                           & 119, 215       & HVC & 18.0       & $-$2.07, $-$1.93 & $-$2.50, $-$1.83 \\
PKS\,0312$-$770\tablenotemark{a}& 70, 280       & HVC & 18.0       & $-$4.08, $-$3.62 & $-$2.50, $-$0.22 \\
                           & 280, 372       & HVC & 18.0       & $-$2.93, $-$2.82 & $-$2.50, $-$1.48 \\
PKS\,0405$-$123                & 100, 161       & HVC & 19.5       & \nodata          & \nodata \\
NGC\,1705\tablenotemark{a}   & 100, 148       & HVC & 19.5       & \nodata          & \nodata \\
                           & 220, 367       & HVC & 19.5       & $-$3.50, $-$3.32 & $-$1.00, $-$0.61 \\
HS0624$+$6907\tablenotemark{a}& $-$150, $-$90 & HVC & 19.2       & \nodata          & \nodata \\
                           & $-$90, $-$60   & IVC & 19.6       & $-$3.42, $-$3.06 & $-$2.5, $-$0.87 \\ 
                           & 100, 135       & HVC & 19.6       & \nodata          & \nodata \\
PG\,0953$+$415\tablenotemark{a}& $-$171, $-$130 & HVC & 19.3       & \nodata          & \nodata \\
                           & $-$130, $-$100 & HVC & 19.3       & \nodata          & \nodata \\
                           & 85, 165        & HVC & 19.3       & \nodata          & \nodata \\
                           & 165, 195       & HVC & 19.3       & \nodata          & \nodata \\
Ton\,28                      & 97, 183        & HVC & 19.5       & \nodata          & \nodata \\
3C\,249                      & $-$90, $-$30   & IVC & 19.7       & $-$3.42, $-$3.19 & $-$0.97, $-$0.46 \\
NGC\,3516\tablenotemark{a}   & $-$187, $-$115 & HVC & 19.4, 19.5 & $-$2.73, $-$2.39 & $-$2.50, $-$2.18 \\
                           & $-$90, $-$30   & IVC & 19.8       & $-$3.61, $-$2.19 & $-$2.50, $-$0.00 \\
PG\,1116$+$215\tablenotemark{a}& $-$90, $-$25   & IVC & 19.8       & $-$3.36, $-$3.10 & $-$1.23, $-$0.62 \\
                           & 25, 90         & IVC & 19.8       & \nodata          & \nodata \\
                           & 90, 125        & HVC & 19.1       & \nodata          & \nodata \\
                           & 130, 220       & HVC & 19.5       & $-$2.83, $-$2.76 & $-$2.50, $-$1.82 \\
NGC\,3783                    & 100, 153       & HVC & 18.3       & \nodata          & \nodata \\
                           & 153, 203       & HVC & 18.0       & \nodata          & \nodata \\
                           & 203, 288       & HVC & 19.2       & \nodata          & \nodata \\
NGC\,4051                    & $-$90, $-$30   & IVC & 19.6, 19.7 & $-$3.49, $-$2.30 & $-$2.27, $-$0.00 \\
NGC\,4151                    & $-$90, $-$50   & IVC & 19.4       & $-$3.79, $-$2.50 & $-$2.50, $-$0.00 \\
PG\,1211$+$143\tablenotemark{a}& 47, 90         & IVC & 19.5       & \nodata          & \nodata \\
                           & 155, 208       & HVC & 19.5       & \nodata          & \nodata \\
                           & 260, 293       & HVC & 19.5       & \nodata          & \nodata \\
PG\,1216$+$069                 & 165, 234       & HVC & 19.5       & \nodata          & \nodata \\
                           & 234, 302       & HVC & 19.5       & $-$3.01, $-$2.67 & $-$2.50, $-$2.16 \\
Mrk\,205\tablenotemark{a}    & $-$225, $-$175 & HVC & 19.4, 19.5 & \nodata          & \nodata \\
                           & $-$175, $-$110 & HVC & 18.1       & $-$3.15, $-$3.05 & $-$2.50, $-$1.91 \\
3C\,273                      & 10, 70         & IVC & 19.3       & $-$3.27, $-$3.08 & $-$1.59, $-$0.82 \\
Q\,1230$+$0115\tablenotemark{a}& 95, 152        & HVC & 19.6       & \nodata          & \nodata \\
                           & 285, 318       & HVC & 19.6       & \nodata          & \nodata \\
NGC\,4593\tablenotemark{a}   & 20, 75         & IVC & 19.1, 19.2 & $-$3.24, $-$2.82 & $-$2.50, $-$1.07 \\
                           & 88, 118        & HVC & 19.5       & \nodata          & \nodata \\
                           & 258, 291       & HVC & 19.5       & \nodata          & \nodata \\
PG\,1259$+$593                 & $-$180, $-$105 & HVC & 19.4       & $-$3.47, $-$3.24 & $-$1.83, $-$0.94 \\
                           & $-$80,  $-$30  & IVC & 19.5       & $-$3.41, $-$2.82 & $-$1.15, $-$0.00 \\
PKS\,1302$-$102              & 200, 340       & HVC & $<$19.56   & \nodata          & \nodata \\
Mrk\,279                     & $-$226, $-$115 & HVC & 19.2, 19.3 & \nodata          & \nodata \\
                           & $-$105, $-$55  & IVC & 19.7       & $-$3.50, $-$3.13 & $-$1.08, $-$0.22 \\
NGC\,5548                    & $-$80,  $-$25  & IVC & 19.1, 19.2 & $-$3.80, $-$2.45 & $-$2.50, $-$0.00 \\
Mrk\,1383                    & $-$90, $-$55   & IVC & 19.4       & $-$3.03, $-$2.67 & $-$2.50, $-$1.68 \\
PG\,1444+407\tablenotemark{a}& $-$164, $-$109 & HVC & 19.5       & \nodata          & \nodata \\
                           & $-$109, $-$60  & IVC & 19.5       & \nodata          & \nodata \\
                           & $-$60, $-$20   & IVC & 19.4       & $-$3.71, $-$2.39 & $-$2.50, $-$0.00 \\
Mrk\,876\tablenotemark{a}    & $-$210, $-$155 & HVC & 18.9, 19.0 & $-$3.42, $-$3.31 & $-$2.32, $-$1.10 \\
                           & $-$155, $-$90  & HVC & 19.0       & $-$3.34, $-$3.26 & $-$1.52, $-$0.53 \\
HS1700$+$6416                & $-$197, $-$85  & HVC & 19.2, 19.3 & $-$3.42, $-$3.20 & $-$1.28, $-$0.52 \\
3C\,351\tablenotemark{a}     & $-$221, $-$150 & HVC & 19.6       & \nodata          & \nodata \\
                           & $-$150, $-$107 & HVC & 19.6       & $-$3.11, $-$3.08 & $-$2.50, $-$1.98  \\
                           & $-$107, $-$45  & IVC & 19.6       & $-$3.19, $-$2.99 & $-$2.50, $-$1.24 \\
H\,1821$+$643                  & $-$165, $-$75  & HVC & 19.6, 19.7 & $-$3.00, $-$2.78 & $-$2.50, $-$1.09 \\
Mrk\,509\tablenotemark{a}    & $-$338, $-$256 & HVC & 19.6, 19.7 & \nodata          & \nodata \\
                           & $-$256, $-$213 & HVC & 19.6, 19.7 & \nodata          & \nodata \\
                           & 35, 90         & IVC & 19.5, 19.6 & $-$3.31, $-$3.19 & $-$0.92, $-$0.42 \\
                           & 100, 147       & HVC & 19.6, 19.7 & \nodata          & \nodata \\
IR\,2121$-$1757                & 30, 80         & IVC & 19.7, 19.8 & $-$3.42, $-$3.20 & $-$1.28, $-$0.52 \\
PHL\,1811\tablenotemark{a}   & $-$385, $-$328 & HVC & 19.5       & \nodata          & \nodata \\
                           & $-$309, $-$237 & HVC & 19.5, 19.5 & \nodata          & \nodata \\
                           & $-$237, $-$180 & HVC & 19.5       & $-$2.79, $-$2.54 & $-$2.50, $-$2.02 \\
                           & $-$180, $-$128 & HVC & 19.5       & \nodata          & \nodata \\
PKS\,2155-304\tablenotemark{a}& $-$309, $-$215& HVC & 19.5       & \nodata          & \nodata \\
                           & $-$195, $-$91  & HVC & 19.5       & \nodata          & \nodata \\
UGC\,1216\tablenotemark{a}  & $-$477, $-$381 & HVC & 19.7       & $-$2.78, $-$2.73 & $-$2.50, $-$2.45 \\
                           & $-$372, $-$324 & HVC & 19.7       & \nodata          & \nodata \\
                           & $-$172, $-$135 & HVC & 19.7       & \nodata          & \nodata \\
                           & $-$90, $-$35   & IVC & 19.6       & $-$3.50, $-$2.99 & $-$1.22, $-$0.15 \\
NGC\,7469\tablenotemark{a}   & $-$410, $-$270 & HVC & 19.5       & $-$2.68, $-$2.56 & $-$2.50, $-$1.83 \\
                           & $-$270, $-$237 & HVC & 19.5       & \nodata          & \nodata \\
                           & $-$190, $-$114 & HVC & 19.5       & \nodata          & \nodata \\
                           & $-$60, $-$15   & IVC & 19.4       & $-$2.98, $-$2.68 & $-$2.17, $-$1.47 \\
\enddata
\tablenotetext{a}{Columns list ranges of model parameters, log\,\NHI, $\log\, U$, and $\log\, (Z/Z_\sun)$.  There are multiple velocity ranges for these sight lines, each treated as a separate HVC or IVC.}
\end{deluxetable}

\end{document}